\documentclass[a4paper,12pt]{article}
\usepackage{jheppub}

\pdfoutput=1

\usepackage[table]{xcolor}
\usepackage{amsmath,array,amsfonts,graphicx,wrapfig,tabstackengine,lscape,float,mathtools,multirow,longtable,setspace}
\stackMath

\usepackage[american]{babel}
\usepackage{physics}
\usepackage{cleveref}

\usepackage{subcaption}
\usepackage{cancel}

\newcommand{\be}{\begin{equation}}
\newcommand{\ee}{\end{equation}}
\newcommand{\beq}{\begin{equation}}
\newcommand{\beql}[1]{\begin{equation}\label{#1}}
\newcommand{\eeq}{\end{equation}}
\newcommand{\ba}{\begin{array}}
\newcommand{\ea}{\end{array}}
\newcommand{\bea}{\begin{eqnarray}}
\newcommand{\beal}[1]{\begin{eqnarray}\label{#1}}
\newcommand{\eea}{\end{eqnarray}}
\newcommand{\ben}{\begin{enumerate}}
\newcommand{\een}{\end{enumerate}}
\newcommand{\bean}{\begin{eqnarray*}}
\newcommand{\eean}{\end{eqnarray*}}
\newcommand{\eref}[1]{(\ref{#1})}
\newcommand{\sref}[1]{\S\ref{#1}}

\newcommand{\fref}[1]{Figure \ref{#1}}
\newcommand{\btab}[1]{\begin{tabular}{#1}}
\newcommand{\etab}{\end{tabular}}

\def \Black {\pdfsetcolor{0 0 0 1}}

\newcommand{\comment}[1]{}

\newcommand{\qed}{\nobreak \ifvmode \relax \else
      \ifdim\lastskip<1.5em \hskip-\lastskip
      \hskip1.5em plus0em minus0.5em \fi \nobreak
      \vrule height0.75em width0.5em depth0.25em\fi}

\definecolor{darkspringgreen}{rgb}{0.09, 0.45, 0.27}
\definecolor{forestgreen}{rgb}{0.13, 0.55, 0.13}
\definecolor{blue2}{RGB}{65,128,255}
\definecolor{yellow2}{rgb}{0.98, 0.80, 0.20}

\newcommand{\wb}{\overline}

\usepackage{youngtab}
\usepackage{array}
\usepackage{tikz}
\usepackage{tikz-3dplot}
\tikzset{cross/.style={cross out, draw=black, minimum size=2*(#1-\pgflinewidth), inner sep=0pt, outer sep=0pt},
cross/.default={1pt}}

\usepackage{pifont}
\newcommand{\cmark}{\ding{51}}%
\newcommand{\xmark}{\ding{55}}%

\newcolumntype{C}[1]{>{\centering\let\newline\\\arraybackslash\hspace{0pt}}m{#1}}

\newcommand{\drawsquare}[2]{\hbox{%
\rule{#2pt}{#1pt}\hskip-#2pt
\rule{#1pt}{#2pt}\hskip-#1pt
\rule[#1pt]{#1pt}{#2pt}}\rule[#1pt]{#2pt}{#2pt}\hskip-#2pt
\rule{#2pt}{#1pt}}

\newcommand{\symm}{~\raisebox{-.5pt}{\drawsquare{6.5}{0.4}}\hskip-0.4pt%
        \raisebox{-.5pt}{\drawsquare{6.5}{0.4}}~}

\newcommand{\beas}{\begin{equation} \begin{aligned}} \newcommand{\eeas}{\end{aligned} \end{equation}}

\title{Dimers, Orientifolds and Stability of Supersymmetry Breaking Vacua} 

\author[a]{Riccardo Argurio}
\author[b]{Matteo Bertolini}
\author[c,d,e]{Sebasti\'an Franco}
\author[a]{Eduardo Garc\'{\i}a-Valdecasas}
\author[b]{Shani Meynet}
\author[a]{Antoine Pasternak}
\author[f]{Valdo Tatitscheff}

\affiliation[a]{Physique Th\'eorique et Math\'ematique and International Solvay Institutes \\ Universit\'e Libre de Bruxelles; C.P. 231, 1050 Brussels, Belgium}

\affiliation[b]{SISSA and INFN, Via Bonomea 265; I 34136 Trieste, Italy}

\affiliation[c]{Physics Department, The City College of the CUNY \\ 160 Convent Avenue, New York, NY 10031, USA}

\affiliation[d]{Physics Program and $^e$Initiative for the Theoretical Sciences \\
The Graduate School and University Center, The City University of New York  \\
365 Fifth Avenue, New York NY 10016, USA}

\affiliation[f]{IRMA, UMR 7501, Universit\'e de Strasbourg et CNRS \\ 
7 rue Ren\'e Descartes 67000 Strasbourg, France}

\emailAdd{rargurio@ulb.ac.be}
\emailAdd{bertmat@sissa.it}
\emailAdd{sfranco@ccny.cuny.edu}
\emailAdd{eduardo.garcia.valdecasas@gmail.com}
\emailAdd{smeynet@sissa.it}
\emailAdd{antoine.pasternak@ulb.ac.be}
\emailAdd{valdotatitscheff@gmail.com}

\abstract{We study (orientifolded) toric Calabi-Yau singularities in search for D-brane configurations which lead to dynamical supersymmetry breaking at low energy. By exploiting dimer techniques we are able to determine that while most realizations lead to a Coulomb branch instability, a rather specific construction admits a fully stable supersymmetry breaking vacuum. We describe the geometric structure that a singularity should have in order to host such a construction, and present its simplest example, the Octagon. 
}
\begin{document}

\maketitle

\section{Introduction}

Since the early days of the AdS/CFT correspondence \cite{Maldacena:1997re,Witten:1998qj,Gubser:1998bc} 
and its non-conformal extensions the possibility of describing, holographically, supersymmetric gauge theories enjoying different IR behaviors has been thoroughly investigated. This has been a rich and lively arena in the field and remarkable results have been obtained in the last two decades. 

While conformal phases, confinement, generation of a mass gap, Coulomb and Higgs-like branches and more generally any supersymmetry preserving dynamics were reproduced in a plethora of examples, not surprisingly (dynamical) supersymmetry breaking has proven to be much harder to achieve. Known examples describe supersymmetry breaking into metastable vacua (see \cite{Kachru:2002gs,Franco:2006es,Argurio:2006ny,Argurio:2007qk} and many other constructions thereafter) or runaway behavior, where the theory breaks supersymmetry dynamically but it does not enjoy a vacuum at finite distance in the space of field VEVs \cite{Berenstein:2005xa,Franco:2005zu,Bertolini:2005di,Intriligator:2005aw}, very much like massless SQCD with a small number of flavors. No models were known, until recently, that enjoy dynamical supersymmetry breaking (DSB) into stable vacua. 

The difficulty in finding such models can suggest that DSB into stable vacua might not be a possibility in D-brane constructions and, more generally, in string theory as well. On the other hand, finding models of this kind could be of great relevance both in the context of the gauge/gravity duality and, even more interestingly, in string compactifications. In this latter setup 
they could be used for model building in GKP-like constructions
\cite{Giddings:2001yu}. Eventually, they might also have an impact on the swampland program \cite{Vafa:2005ui,Brennan:2017rbf,Palti:2019pca} and recent related conjectures such as \cite{Buratti:2018onj}. 

Recently, a series of papers renewed the interest in models of D-branes at Calabi-Yau (CY) singularities leading to dynamical supersymmetry breaking. This originated from \cite{Franco:2007ii} where an existence proof for a possibly stable DSB model obtained by considering fractional branes at orientifold singularities was given. These results were generalized in \cite{Argurio:2019eqb}, where it was shown that a large class of orientifolds admit fractional D-brane configurations realizing some of the most popular and simple DSB models, namely the incalculable $SU(5)$  \cite{Affleck:1983vc}  and/or 3-2 \cite{Affleck:1984xz} models. 

In this same work \cite{Argurio:2019eqb}, however, by generalizing previous results of \cite{Buratti:2018onj}, it was shown that in the decoupling limit \cite{Maldacena:1997re}, in which the DSB fractional D-brane bound state becomes part of a UV complete large $N$ D-brane model and gravity is decoupled, all models display an instability. This instability turned out to have a common, model-independent geometric origin in terms of $\mathcal{N}=2$ fractional branes probing the singularity.\footnote{${\cal N} = 2$ fractional  branes arise whenever a Calabi-Yau singularity can be partially resolved to display, locally, a non-isolated $\mathbb{C}^2/\mathbb{Z}_n$ singularity
and a Coulomb-like branch associated to it.} More drastically, a no-go theorem was proven in \cite{Argurio:2019eqb} which implies that whenever $\mathcal{N}=2$ classical flat directions exist at a singularity which admits such DSB models, the quantum behavior of the latter is such that the flat directions are tilted and supersymmetry preserving vacua exist. 

An obvious way to circumvent this no-go theorem and avoid the unwanted slide towards supersymmetric vacua is to look at singularities free of $\mathcal{N}=2$ fractional branes to start with, and see whether stable DSB models of the type above can be engineered there. Or, alternatively, a stronger no-go theorem should exist which excludes such a possibility altogether. 

This is what we will be concerned with in this paper.\footnote{In the same vein as in \cite{Franco:2007ii,Buratti:2018onj,Argurio:2019eqb}, we will not consider configurations where non-compact flavor branes are added. We note that metastable DSB can be engineered in this way \cite{Franco:2006es}, and further investigating if stable DSB is possible in these constructions is an interesting problem that we do not address here.} More precisely, our main goal will be to answer the following question: 

\begin{itemize}
\item Is it possible to get a DSB model, more specifically the $SU(5)$ or the 3-2 models, from D-branes at  a Calabi-Yau singularity which is free of any (known) instability? 
\end{itemize}

Quite surprisingly, the answer will be affirmative! We will first carry out a comprehensive investigation that shows that in the minimal realizations of the $SU(5)$ and 3-2 models at orientifolds of singularities, the instability associated to $\mathcal{N}=2$ fractional branes is unavoidable. Remarkably, this result ties the ability to engineer these models to basic geometric features of the underlying singularity: the presence of non-isolated $\mathbb{C}^2/\mathbb{Z}_n$ singularities. This is yet another example of the connection between geometry and features or dynamics of the corresponding quantum field theories, such as e.g. confinement and complex deformations \cite{Klebanov:2000hb} or runaway DSB and the absence of complex deformations \cite{Berenstein:2005xa,Franco:2005zu}. These general results will then guide our search of models without instabilities. We will show that a simple variant of the $SU(5)$ model, that we dub twin $SU(5)$, can be realized by D-branes at an orientifold of a toric CY, the Octagon, which lacks non-isolated $\mathbb{C}^2/\mathbb{Z}_n$ singularities and, as such, is free of the aforementioned decay channel, and stable. Our analysis, which is done exploiting dimers techniques \cite{Franco:2005rj,Franco:2005sm}, relies also on results obtained in \cite{Argurio:2020dko}, where a thorough investigation of consistent, anomaly free, D-brane models at orientifold singularities has been performed.

While we do not prove nor exclude the existence of other, more involved models sharing the same properties, the example we provide shows that stable DSB can be engineered by brane configurations at CY singularities. Given the implications that this might have in different contexts, including improvements in our understanding of the string landscape and the swampland, it is worth investigating these D-brane constructions further. On a more technical side, the results presented here as well as in \cite{Argurio:2020dko} show, once again, the power of dimer techniques in understanding the properties of D-branes and more generally string theory at CY singularities. 

The rest of the paper is organized as follows. In \sref{rev} we review the basic elements of the $SU(5)$ and 3-2 DSB models and the decay mechanism that afflicts all known D-brane constructions which were found to realize these models in string theory, so far. In the two subsequent sections, \sref{su5} and \sref{32}, we discuss the possibility to embed the $SU(5)$ and the 3-2 models, or any simple variant, into D-brane models at orientifolds singularities which,  locally, do not display any non-isolated $\mathbb{C}^2/\mathbb{Z}_n$ singularity. This will single out the twin $SU(5)$ as a candidate for a fully stable DSB model. In \sref{Octagon} we show that such local construction can be embedded in a fully consistent (orientifolded) dimer, the Octagon, and present the UV complete D-brane gauge theory associated to it. A thorough analysis of the Octagon, including a discussion of its stability properties, can be found in the companion paper \cite{Argurio:2020dkg}. Appendix \sref{dimers} contains a review of dimer models and some technical details which for the sake of the presentation are not included in the main body of the paper. Finally, in Appendix \sref{appendix_ACC_32} some details are given on anomaly cancellation conditions for the models discussed in \sref{32}.

\section{Review of Background Material}
\label{rev}

In this section we go over some known results which will be needed in the following. We will first review the basic structure of the two dynamical supersymmetry breaking models we will be concerned with, the so-called $SU(5)$ \cite{Affleck:1983vc} and  3-2 models \cite{Affleck:1984xz}. Then, we review the mechanism responsible for the instability which afflicts all D-brane configurations realizing the aforementioned DSB models in any of the string theory constructions presented in \cite{Buratti:2018onj,Argurio:2019eqb}. Finally, we review the characterization of $\mathcal{N}=2$ fractional branes according to the general classification of \cite{Franco:2005zu} and discuss their description in terms of dimers.

\subsection{DSB Models}

The $SU(5)$ model is the prototype of a class of models which are believed to break supersymmetry dynamically \cite{Affleck:1983vc}. This is a supersymmetric gauge theory with gauge group $SU(5)$ and one GUT-like chiral family $(\bf{10} \oplus \bar{\bf{5}})$, namely one chiral superfield transforming in the antisymmetric representation and one chiral superfield transforming in the antifundamental representation of $SU(5)$. The theory does not admit any classical flat direction since no gauge invariants can be written and enjoys a $G_F=U(1) \times U(1)_R$ non-anomalous global symmetry. An argument based on the impossibility to satisfy 't~Hooft anomaly matching conditions for $G_F$ suggests that supersymmetry is broken due to strong coupling effects.

The $SU(5)$ model admits several generalizations. The only one which will be realized in our setups is a rather trivial one, where a possibly complicated supersymmetric gauge theory reduces, at low energy, to several decoupled $SU(5)$ models which independently break supersymmetry.

The 3-2 model \cite{Affleck:1984xz} is the simplest representative of the so-called calculable models, which have the property to break supersymmetry dynamically while allowing a region in the parameter space in which the low energy effective theory is completely calculable, in the sense that both the effective superpotential and the K\"ahler potential are under control. The 3-2 model consists of a supersymmetric gauge theory with gauge group $SU(3) \times SU(2)$ and one Standard Model-like family described by chiral superfields transforming as $(\bf{3},\bf{2}) \oplus (\bar{\bf{3}},\bf{1})\oplus (\bar{\bf{3}},\bf{1})\oplus (\bf{1},\bf{2})$, respectively, and a cubic superpotential. Classically, the theory admits an isolated supersymmetric vacuum. However, quantum mechanically a non-perturbative superpotential is generated which makes it impossible to satisfy both the $F$- and $D$-term conditions and no supersymmetric vacuum can be found. The minimum of the scalar potential breaks supersymmetry, the vacuum energy being related to the dynamical scales of the strongly coupled gauge groups. 

As the $SU(5)$ model, the 3-2 model admits several generalizations, of which we will realize only the possibility of multiple, decoupled supersymmetry breaking sectors at low energy.\footnote{Both the $SU(5)$ and the 3-2 models have higher rank generalizations. As for other existing field theoretic models of stable DSB, they seem more difficult to realize through brane constructions, though they certainly deserve a closer look.}

\subsection{$\mathcal{N}=2$ Fractional Brane Decay}
\label{n2fd}

In \cite{Argurio:2019eqb}, generalizing previous results of \cite{Buratti:2018onj}, several D-brane models at orientifold Calabi-Yau singularities realizing either the $SU(5)$ or 3-2 models at low energy were found. They can be engineered by bound states of fractional D3-branes which can arise at the end of complicated RG-flows (often described by a duality cascade \cite{Klebanov:2000hb}) or on the $\mathcal{N}=4$ Coulomb branch of regular  D3-branes, depending on the singularity structure.
 
A common feature of {\it all} Calabi-Yau singularities in which the above DSB models were found is that they admit  $\mathcal{N}=2$ fractional branes. The latter are related to a partial resolution of the  singularity displaying a non-isolated $\mathbb{C}^2/\mathbb{Z}_n$ singularity and an $\mathcal{N}=2$ Coulomb branch associated to it (further details on this class of fractional branes are provided in the coming section). In \cite{Argurio:2019eqb} it was proven in full generality that whenever such $\mathcal{N}=2$ classical flat directions exist at a singularity which admits the aforementioned DSB models, the quantum behavior of the latter tilts the flat directions towards a supersymmetry preserving vacuum. Hence, the supersymmetry breaking vacuum is destabilized and ceases to exist. This novel decay mechanism, which remarkably has an elegant geometric origin, was originally uncovered in \cite{Buratti:2018onj}.

The basic dynamical mechanism describing such instability goes as follows. In the decoupling limit \cite{Maldacena:1997re}, the DSB model emerges as a vacuum configuration of a (possibly intricate) system of regular and fractional D-branes, with the vacuum energy depending on the VEVs of the scalar fields. The $\mathcal{N}=4$ Coulomb branch is parametrized by regular branes. If an $\mathcal{N}=2$ fractional brane direction exists, there is in addition an $\mathcal{N}=2$ Coulomb branch. By scale matching, one can show that the energy of the supersymmetry breaking vacuum is related to the strong coupling scale $\Lambda$ of the $SU(5)$ or $SU(3)\times SU(2)$ gauge groups as follows\footnote{In the 3-2 model, $\Lambda$ refers to the scale of the gauge group factor, either $SU(3)$ or $SU(2)$, that dominates the dynamics.}
\be
\label{eqdec1}
E_{vac} = \left(\frac{v'}{v}\right)^\alpha \Lambda~~,~~\alpha  \in \mathbb{R}~,
\ee
where the exponent $\alpha$ is given by a ratio of beta functions and $v$ and $v'$ are the VEVs on the Coulomb branches associated to the $\mathcal{N}=2$ fractional brane and its complement, respectively. Fractional branes are defined modulo regular branes, so that a fractional brane and its complement combine into a regular brane.  The case $v=v'$ then corresponds to the $\mathcal{N}=4$ Coulomb branch.

From \eref{eqdec1}, it follows that on the $\mathcal{N}=4$ Coulomb branch the vacuum energy equals $\Lambda$ and the supersymmetry breaking vacuum is hence preserved. 
On the $\mathcal{N}=2$ Coulomb branch, instead, where $v \not =v'$, the vacuum energy relaxes to 0, with a moduli space parametrized by $v$ at $v'=0$ or vice-versa, depending on the sign of $\alpha$. Geometrically, this corresponds to a supersymmetric configuration described by the $\mathcal{N}=2$ fractional branes associated to $v$ located at a finite distance along the non-isolated $\mathbb{C}^2/\mathbb{Z}_n$ singularity describing its Coulomb branch, and their complement  sitting at the origin. 

The only possibility for evading this decay mechanism of the supersymmetry breaking vacuum is that $\alpha=0$. Using some basic properties of Calabi-Yau's and the fact that fractional branes are described by a non-conformal field theory at low energy, one can easily show that $\alpha \not =0$ \cite{Argurio:2019eqb}. The upshot is that all DSB D-brane models constructed in \cite{Buratti:2018onj,Argurio:2019eqb} are actually unstable since, as anticipated, all of them admit $\mathcal{N}=2$ fractional branes. At most they can be metastable.

Let us schematically discuss what occurs to the gauge theory in this process.  We denote $G_{\cancel{SUSY}}$ the SUSY breaking model, namely its gauge group (and possibly flavor group), matter fields and interactions. When $N$ regular D3-branes are added, the SUSY breaking sector extends to 
\beq
G_{\cancel{SUSY}} \ \ \ \xrightarrow[\mbox{\footnotesize{branes}}]{\mbox{\footnotesize{   + $N$ regular   }}} \ \ \ G_{\cancel{SUSY}+N} \times G'_N~,
\eeq
where $G_{\cancel{SUSY}+N}$ indicates that the ranks of the gauge and flavor groups are increased by $N$. $G'_N$ denotes the theory associated to the complement. The subindex indicates that all gauge groups in this sector have rank $N$. In addition, there is matter connecting the $G_{\cancel{SUSY}+N}$ and $G'_N$ sectors. Along the $\mathcal{N}=2$ Coulomb branch, the theory is higgsed down to 
\beq
G_{\cancel{SUSY}+N} \times G'_N \ \ \ \xrightarrow{\mbox{\footnotesize{   $v \neq v'$   }}} \ \ \ G_{\cancel{SUSY}} \times G'_N ~,
\label{2_sectors_higgsed}
\eeq
We are left precisely with the original SUSY breaking theory of interest, but now coupled to $G'_N$. This extension of the theory spoils supersymmetry breaking.

The only way to avoid this decay mechanism is to look for singularities that admit supersymmetry breaking D-brane configurations {\it and} are free of local $\mathbb{C}^2/\mathbb{Z}_n$ singularities. Whether such geometries exist or not was until now an open question. Answering this question has been one of the main motivations for the present work.

\subsection{$\mathcal{N}=2$ Fractional Branes and Dimers}
\label{section_N=2_dimers}

In order to be self-contained, we now present a quick review of $\mathcal{N}=2$ fractional branes, which play a central role in this paper. A classification of fractional branes based on the IR dynamics of the gauge theories living on them was introduced in \cite{Franco:2005zu}. According to it, fractional branes fall into three classes: deformation, $\mathcal{N}=2$ and DSB fractional branes (we refer to \cite{Franco:2005zu} for a thorough discussion).

These fractional branes support a gauge invariant operator that does not appear in the superpotential. In dimer language, they correspond to a collection of faces forming a stripe, which gives rise to the closed path in the quiver associated to this gauge invariant operator. The VEV of such operator parameterizes a flat direction along which the dynamics reduces to an $\mathcal{N}=2$ theory. 
Geometrically, $\mathcal{N}=2$ fractional branes arise in the case of non-isolated singularities, which have complex curves of singularities passing through the origin. Such fractional branes wrap a 2-cycle collapsed at the singularity, which exists at every point along the curve. In the case of toric geometries, the singularity on the curve is always $\mathbb{C}^2/\mathbb{Z}_n$, $n\geq 2$. Such a singularity translates into edges on the boundary of the toric diagram with $n-1$ internal points. Equivalently, they correspond to $n$ parallel legs in the dual $(p,q)$-web diagram \cite{Aharony:1997ju,Aharony:1997bh,Leung:1997tw}.  Legs of the $(p,q)$-web are in one to on correspondence with {\it zig-zag paths} (ZZP) in the dimer (see Appendix \sref{dimers} for more details). Indeed, the stripe of faces in the dimer describing an $\mathcal{N}=2$ stretches between a pair of ZZP with the same holonomy. Finally, let us emphasize that the previous discussion implies that the gauge theories/dimers associated to toric diagrams without internal points on external edges, i.e.~without non-isolated singularities, do not support $\mathcal{N}=2$ fractional branes.

\begin{figure}[h!]
  \centerline{\includegraphics[width=8cm]{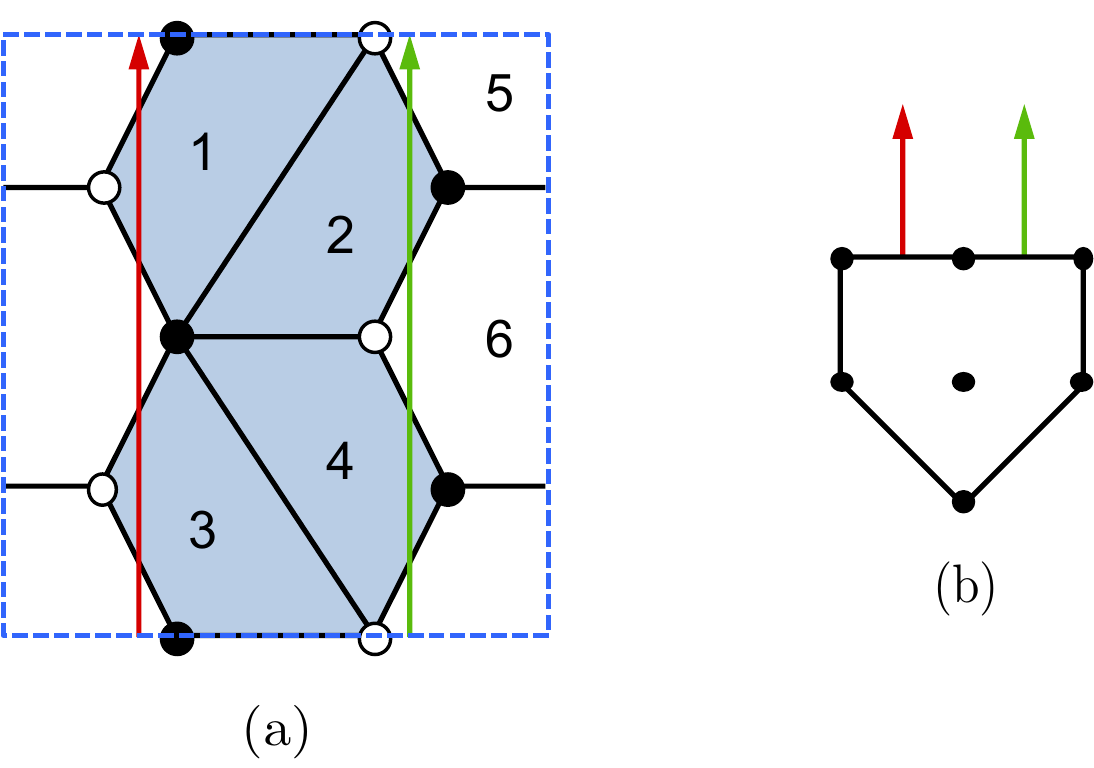}}
    \caption{a) Dimer model for phase $b$ of $PdP_3$. b) Toric diagram for $PdP_3$ showing the two parallel legs of the $(p,q)$-web associated to the ZZP under consideration.}
\label{example_N=2_branes}
\end{figure}

\fref{example_N=2_branes} shows an example, based on the $PdP_3$ geometry \cite{Feng:2002fv,Argurio:2019eqb}, illustrating the ideas presented above. The collection of phases shaded in blue defines an $\mathcal{N}=2$ fractional brane (its complement is obviously also an $\mathcal{N}=2$ fractional brane). These faces stretch between the parallel red and green ZZP.

\section{$\boldsymbol{SU(5)}$ Models}
\label{su5}

Let us first consider the $SU(5)$ model. This theory has an $SU(5)$ gauge group and one GUT-like chiral family ${\tiny \yng(1,1)}\, \oplus \, \tiny\overline{\yng(1)}$\ . The presence of the antisymmetric representation implies that if one wants to engineer such a model by D-branes at a CY singularity, an orientifold projection is necessary. Moreover, one has to consider two gauge groups in order to get the antifundamental representation $\tiny\overline{\yng(1)}$, which can be generated by either an $SU(1)$ or an $SO(1)$ flavor group \cite{Argurio:2019eqb}.  

Using the dimer formalism, there are two classes of orientifolds, depending on whether they have {\it fixed points} or {\it fixed lines} \cite{Franco:2007ii} (see Appendix \ref{dimers} for a short review).\footnote{In this paper we will restrict to the class of orientifolds considered in \cite{Franco:2007ii}, with either fixed points or fixed lines. In principle, $\mathbb{Z}_2$ involutions of the dimer without fixed loci are possible, but they have not been investigated in the literature.} We will analyze them in turn.

\subsection{Fixed Point Orientifolds}

\label{section_SU(5)_fixed_points}

\begin{figure}[h!]
	\centerline{\includegraphics[width=5cm]{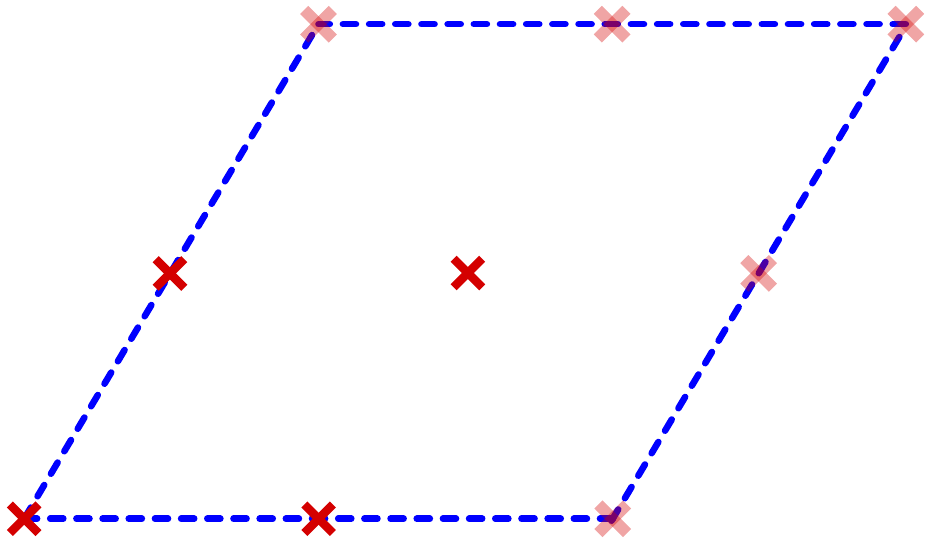}}
	\caption{A schematic representation of a dimer unit cell with  orientifold fixed points. The shaded points are the periodic images of the four basic ones.} 
		\label{opoints}
\end{figure}

Let us remind that fixed point orientifolds are associated to dimers which enjoy a point reflection. It is always possible to choose the unit cell of the dimer in such a way that its corners coincide with a fixed point. Additionally, due to the dimer's toroidal periodicity, there will also be fixed points at the center of the boundaries of the unit cell, and in the center of the unit cell itself, see \fref{opoints}.

As we now review, we not only need a fixed point on one edge of the $SU(5)$ face, but a second fixed point is needed to avoid anomalies in the face providing the (anti)fundamental matter field. 

The first possibility is to directly avoid the anomaly in the flavor group by having it $SO$ or $USp$. $USp$ is ruled out since it would give always an even number of antifundamentals, hence more than one. We are then left with $SO(1)$.

\begin{itemize}

\item \underline{$SO$ flavor group}

\fref{SU5SO} shows the generic structure of a local configuration of a dimer leading to the $SU(5)$ model, including the signs for the two relevant fixed points. The dotted lines and nodes represent a completely general configuration for the rest of the dimer, only constrained by its compatibility with the point reflections. The blue dotted line indicates that it is possible to choose the unit cell such that the two fixed points live on one of the four segments that form its boundary. This comment will be relevant later.

\begin{figure}[h!]
	\centerline{\includegraphics[width=0.47\linewidth]{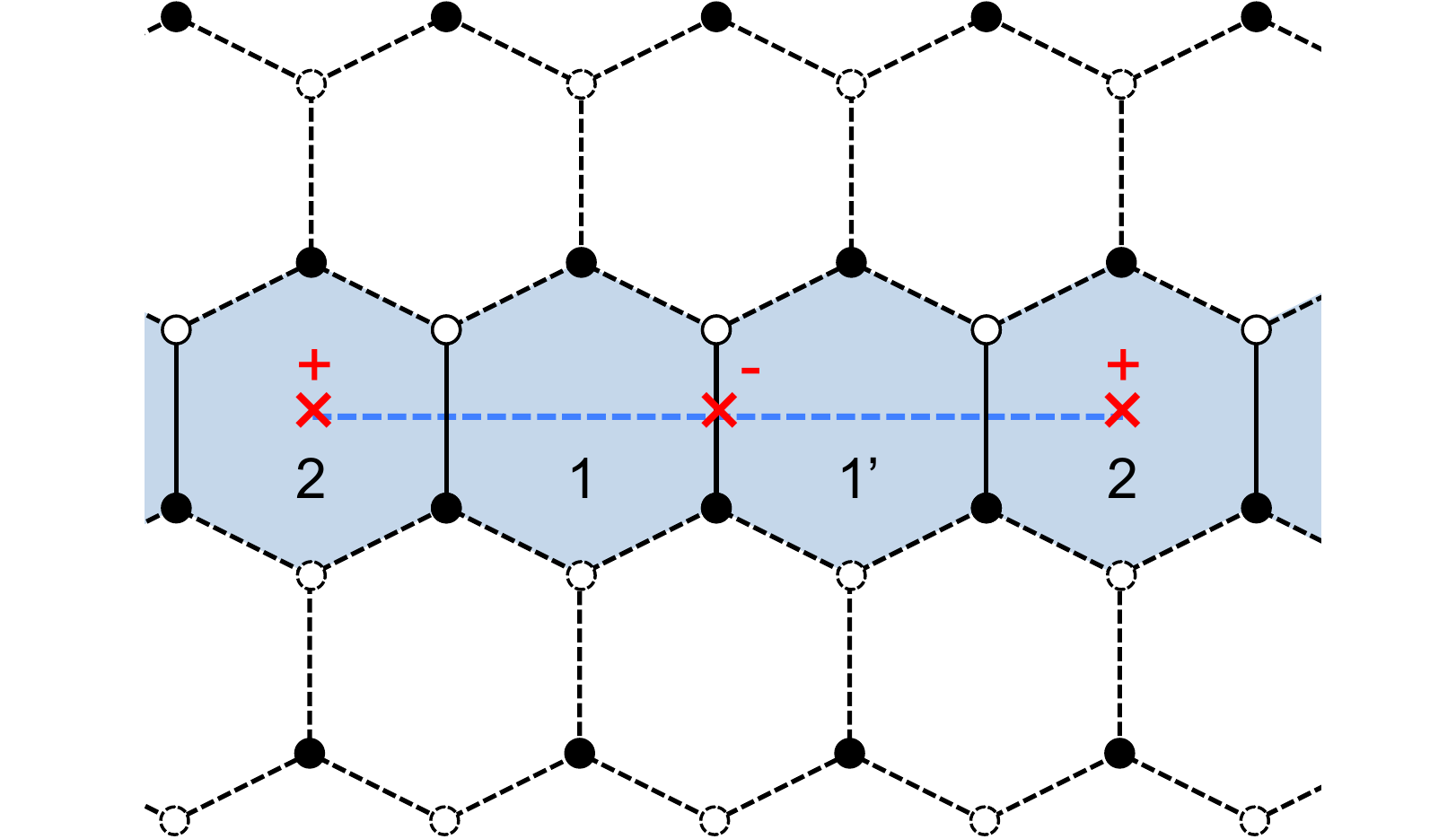}}
	\caption{Fixed point orientifold realizing the $SU(5)$ model with $SO(1)$ flavor group. The dotted part of the graph indicates the rest of the dimer, which is completely general and not necessarily hexagonal as shown.} 
		\label{SU5SO}
\end{figure}

Assigning arbitrary ranks to the gauge groups, $N_i$ for face $i$ in the dimer, the anomaly cancellation conditions (ACC) have a solution in which $N_1= N_{1'}=5$, $N_2=1$ and the rest of the faces are empty.\footnote{Of course whether the ACC of the empty nodes are also satisfied depends on the details of the boundary of the cluster of faces under consideration. This observation also applies to the examples that follow.} This choice leads exactly to the $SU(5)$ model. Face 1 becomes the $SU(5)_1$ gauge group. Since face 2 has a fixed point with a positive sign on top of it, becomes the $SO(1)_2$ flavor group.

\end{itemize}

A second possibility is that the flavor group is of $SU$ type, with its anomaly (when regular branes are added) being canceled by the presence of symmetric matter on a different edge of the face. 

\begin{itemize}

\item \underline{$SU$ flavor group with symmetric}

\fref{SU5SU} shows the  local configuration of a dimer leading to another realization of the $SU(5)$ model in a fixed point orientifold. Once again, the ACC have a solution in which $N_1= N_{1'}=5$, $N_2=N_{2'}=1$ and the rest of the faces are empty. The resulting theory is the $SU(5)$ model, plus a decoupled singlet corresponding to the symmetric associated to the edge between face 2 and its image. 

\begin{figure}[h!]
	\centerline{\includegraphics[width=0.47\linewidth]{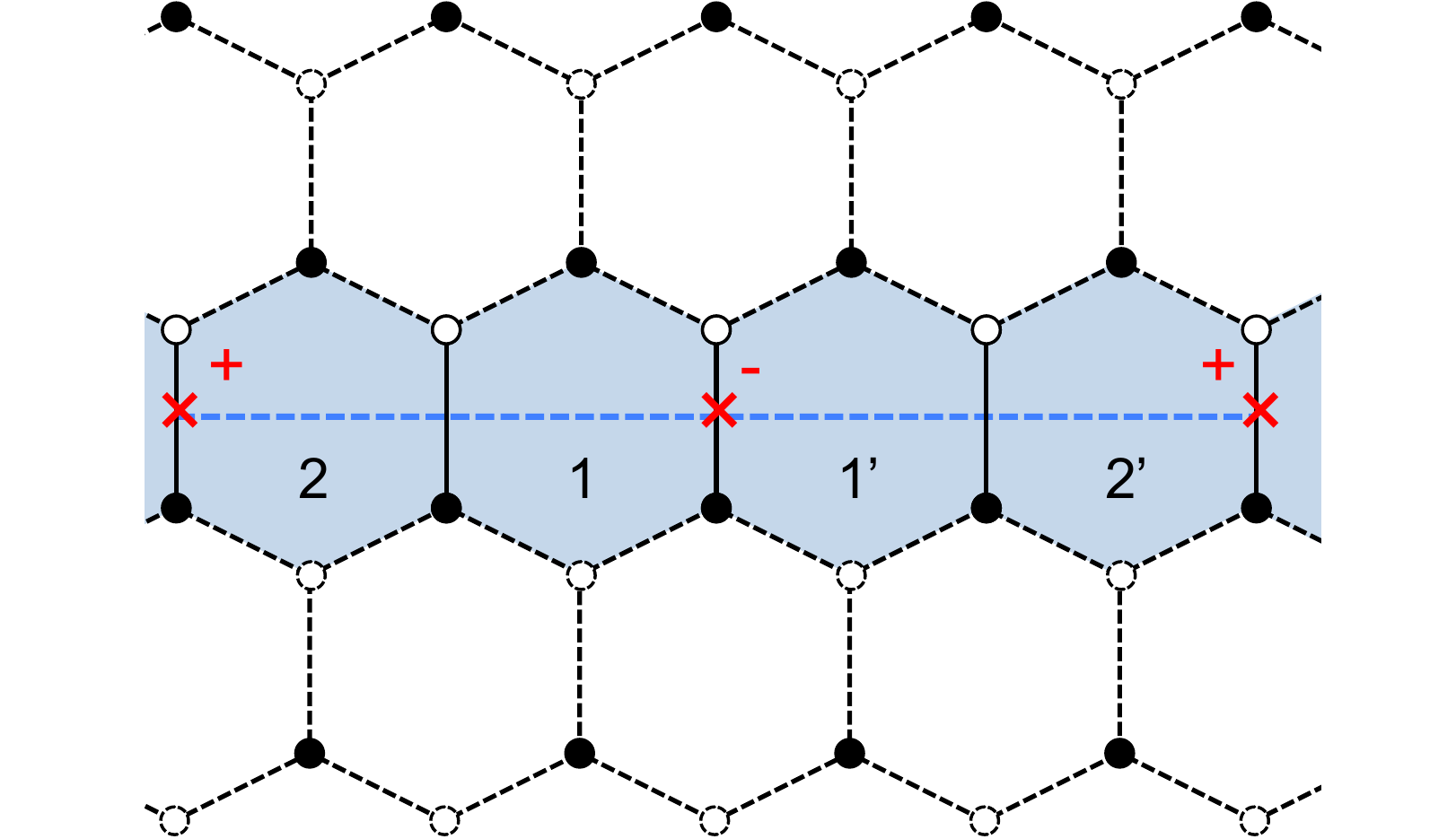}}
	\caption{Fixed point orientifold realizing the $SU(5)$ model with $SU(1)$ flavor group.}
	\label{SU5SU}
\end{figure}

Note that the $SU(1)$ group has no anomaly, but the symmetric is necessary to cancel the anomaly when all the ranks are increased by $N$ (corresponding to the addition of $N$ regular D3-branes which populate the dimer democratically). By construction, the additional (white) faces with rank $N$ will not contribute to the anomaly. In order to cancel the $N+5$ antifundamentals coming from face 1, we need to have a symmetric of $SU(N+1)$ at face 2. It reduces to a decoupled singlet when $N=0$.

\end{itemize}

\begin{figure}[h!]
  \centerline{\includegraphics[width=0.63\linewidth]{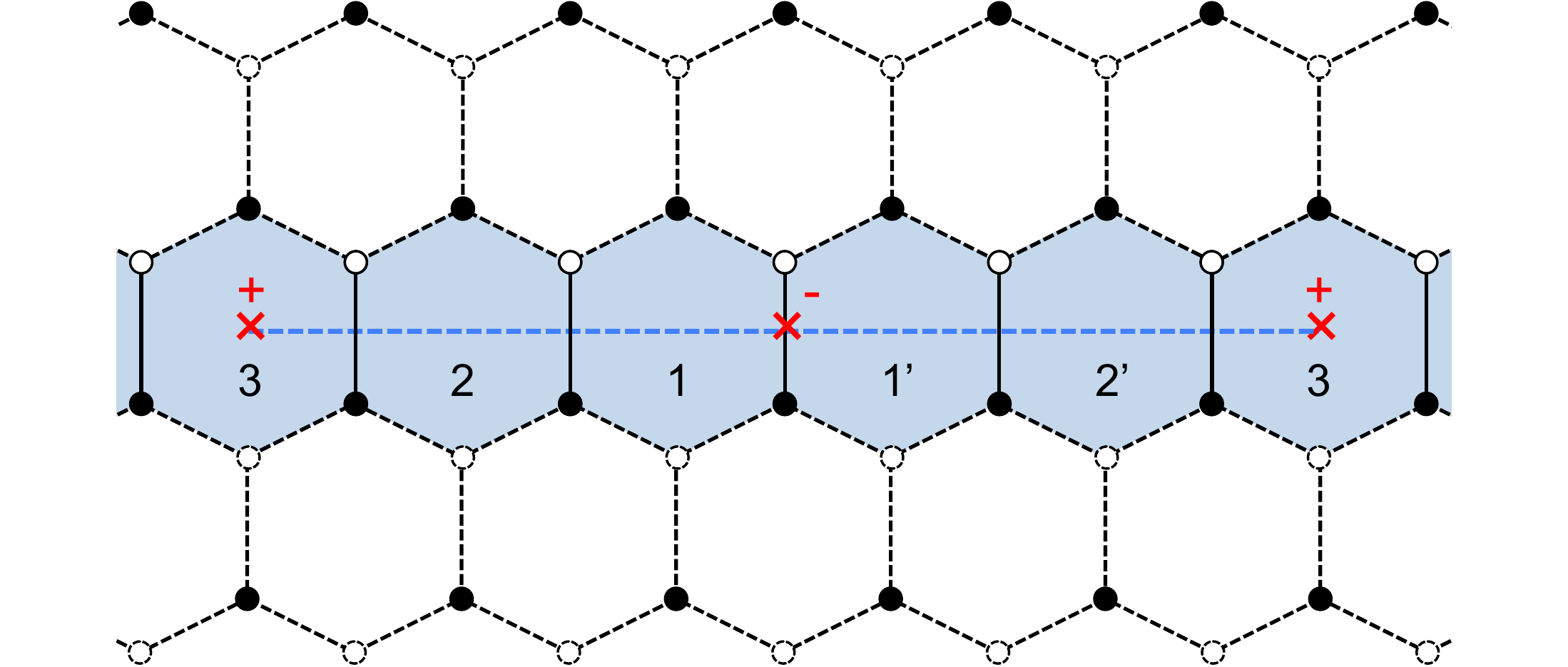}}
    \caption{Fixed point orientifold realizing the $SU(5)_1$ model with $SU(1)_2$ flavor group and an additional $SO(5)_3$ factor. $SO(5)_3$ develops an ADS superpotential and leads to a runaway behavior.}
\label{SU5SUSO5_points}
\end{figure}

A third possibility is that the flavor group is of $SU$ type, and its anomaly (when regular branes are added) is canceled by 5 fundamentals attached to an $SO(5)$ group. This configuration is shown in \fref{SU5SUSO5_points}. The low-energy theory of this configuration is an $SU(5)$ model together with a decoupled $SO(5)$ SQCD with one flavor. The latter theory develops an ADS superpotential \cite{Affleck:1984xz}, so that we have a runaway behavior (on top of the DSB of the $SU(5)$ model), and hence no true vacuum. We thus discard this possibility since it is already unstable at this low-energy field theory level.

A fourth possibility is that the flavor group is again of $SU$ type, but now its anomaly is canceled by the presence of a replica of the $SU(5)$ group with its own antisymmetric. We will call this possibility {\it twin $SU(5)$ model}.

\begin{itemize}
	
\item \underline{$SU$ flavor group with twin $SU(5)$}
	
\fref{twin_SU5_points} shows the  local configuration of a dimer leading to yet another realization of the $SU(5)$ model in a fixed point orientifold. The ACC have a solution in which $N_1=N_{1'}=5$, $N_2=N_{2'}=1$, $N_3=N_{3'}=5$ and the rest of the faces are empty. The resulting theory corresponds to two $SU(5)$ models sharing one and the same $SU(1)$ flavor group which provides their (anti)fundamentals. Since $SU(1)$ is actually empty, and in any case no chiral gauge invariants can be written for each $SU(5)$ model, the twins are effectively decoupled and thus their low-energy dynamics is completely independent. 

\begin{figure}[h!]
  \centerline{\includegraphics[width=0.63\linewidth]{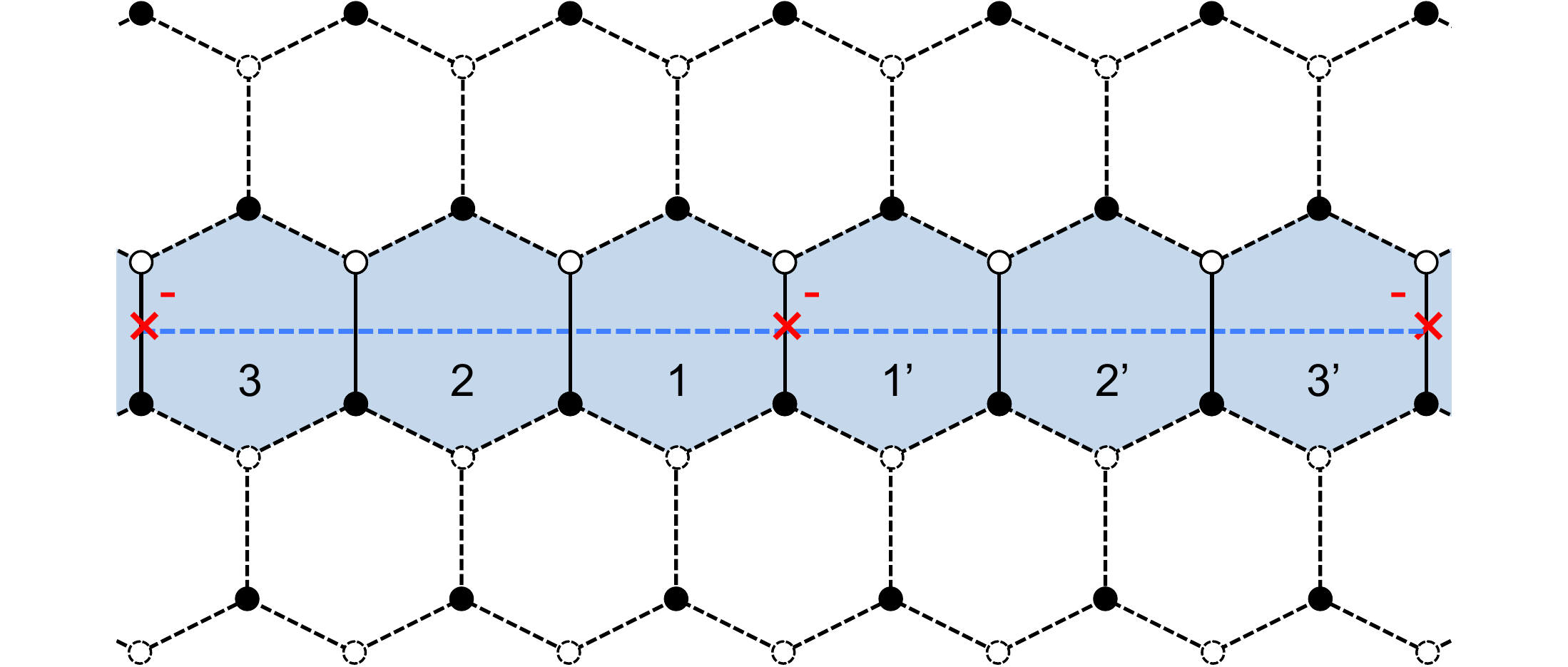}}
    \caption{Fixed point orientifold realizing the twin $SU(5)$ model.}
\label{twin_SU5_points}
\end{figure}
	
\end{itemize}

In principle, we could go on with further possibilities. Indeed, the anomaly of the second $SU(5)$ gauge group at face 3 can be canceled with a fundamental, instead of an antisymmetric. The simplest possibility is that the fundamental is attached to an $SO(1)$ face, however it could also be an $SU(1)$ with a symmetric, or further an $SU(1)$ with 5 antifundamentals given by an $SO(5)$, or another $SU(5)$. The possibilities already discussed above repeat themselves. What is important to notice is that the gauge theory on face 3 would always be an $SU(5)$ with one flavor, hence developing an ADS superpotential and leading to runaway behavior. 

We thus conclude that the only possibilities to engineer an $SU(5)$ model, which is stable at low-energies, in a dimer with fixed points are the three bullets above: $SO$ flavor group, $SU$ flavor group with a symmetric and $SU$ flavor group with twin $SU(5)$. 

An important remark is that in all the examples above the following holds: there can be a long chain of gauge groups to eventually cancel the anomaly of the initial $SU(5)$ gauge group, but it always ends with an orientifold fixed point.\footnote{We are ignoring more ramified possibilities. For instance, for an $SU(1)$ flavor at face 2, we could imagine providing the 5 fundamentals from more than one $SO$ gauge group. That would lead to the need of more than one extra fixed point. The other cases can be treated similarly. Thus a more precise statement is that we always need {\em at least} another fixed point to cancel the anomaly of the $SU(5)$ at face 1.} As a consequence, we do not have to look far in order to identify an $\mathcal{N}=2$ fractional brane in these dimers. Remarkably, in all cases the $SU(5)$ model is fully supported on a set of faces that corresponds to an $\mathcal{N}=2$ fractional brane in the parent (i.e., non-orientifolded) theory. From \fref{SU5SO}, \fref{SU5SU} and \fref{twin_SU5_points} we see that in all cases the $SU(5)$ model indeed lives on a stripe that gives rise to a gauge invariant not contained in the superpotential. The expectation value of such operator parametrizes the corresponding Coulomb branch. 

We conclude that an $SU(5)$ model cannot be obtained for this class of orientifolds if the parent theory does not contain line singularities, i.e.~$\mathcal{N}=2$ fractional branes.\footnote{This result is consistent with an observation made in \cite{Retolaza:2016alb}, namely that singularities with deformation branes are incompatible with point projections.} The previous discussion implies that the no-go theorem in \cite{Argurio:2019eqb} cannot be avoided for this class of orientifolds. 

Let us discuss how the instability explained in \sref{n2fd} is realized in these models in more detail. We start with the model with $SO$ flavor, \fref{SU5SO}. After adding $N$ regular D3-branes, the relevant gauge group becomes
\beq
SU(N+5)_1 \times SO(N+1)_2 \ .
\eeq 
Let us denote
\beq
A={\tiny \yng(1,1)_1} \ \ \ \ , \ \ \ \ \overline{Q}={\tiny (\overline{\yng(1)}_1,\yng(1)_2)}
\eeq
where $A$ corresponds to the edge in the dimer between face 1 and its orientifold image and $\overline{Q}$ corresponds to the edge between faces 1 and 2.  The Coulomb branch is parametrized by the expectation value of the gauge invariant going around the stripe. In principle we can build an $SU(N+5)_1$ gauge invariant as
\beq
\phi^{SO}_{ab}=\overline{Q}^i_a \overline{Q}^j_b A_{ij} ~,
\label{gaugeinvSO}
\eeq
where $i,j$ are fundamental indices of $SU(N+5)_1$ and $a,b$ are fundamental indices of $SO(N+1)_2$.
Note that it is in the antisymmetric representation of $SO(N+1)_2$, hence it does not exist for $N=0$, and it has vanishing trace for $N\geq 1$. 

As discussed in \cite{Argurio:2019eqb}, we actually need to go twice around the stripe in order to have a non-vanishing gauge invariant given by
\beq
\langle \delta^{ac}\delta^{bd} \phi^{SO}_{ab}\phi^{SO}_{cd}\rangle~,
\label{vev_Coulomb_branchSO}
\eeq
parametrizing the Coulomb branch. That the gauge invariant still vanishes automatically for $N=0$, is consistent with the fact that the $SU(5)$ model does not have a moduli space and that the additional regular branes are necessary for the instability.

We now consider the case with $SU$ flavor and a symmetric, \fref{SU5SU}. After adding $N$ regular D3-branes, the gauge group becomes
\beq
SU(N+5)_1 \times SU(N+1)_2 ~.
\eeq 
We denote
\beq
A={\tiny \yng(1,1)_1} \ \ \ \ , \ \ \ \ \overline{Q}={\tiny (\overline{\yng(1)}_1,\yng(1)_2)} \ \ \ \ , \ \ \ \ \overline{S}={\tiny \overline{\yng(2)}_2}
\eeq
where now $\overline{S}$ corresponds to the edge between face 2 and its image under the second fixed point. The $SU(N+5)_1$ gauge invariant is
\beq
\phi^{SU}_{ab}=\overline{Q}^i_a \overline{Q}^j_b A_{ij} ~,
\label{gaugeinvSU}
\eeq
where now $a,b$ are fundamental indices of $SU(N+1)_2$.
It is in the antisymmetric representation of $SU(N+1)_2$, hence again it does not exist for $N=0$, and for $N\geq 1$ it cannot be contracted with $\overline{S}^{ab}$ which is symmetric. 
A non-vanishing gauge invariant is given by
\beq
\langle \overline{S}^{ac}\overline{S}^{bd} \phi^{SU}_{ab}\phi^{SU}_{cd}\rangle~,
\label{vev_Coulomb_branchSU}
\eeq
which now parametrizes the Coulomb branch. The same remarks as in the previous case apply.

Finally, let us discuss the last case of the twin $SU(5)$, where the gauge group becomes
\beq
SU(N+5)_1 \times SU(N+1)_2 \times SU(N+5)_3\ .
\eeq 
We denote
\beq
A={\tiny \yng(1,1)_1} \ \ \ \ , \ \ \ \ \overline{Q}={\tiny (\overline{\yng(1)}_1,\yng(1)_2)} \ \ \ \ , \ \ \ \ \overline{P}={\tiny (\overline{\yng(1)}_2,\yng(1)_3)} \ \ \ \ , \ \ \ \ \overline{A}={\tiny \overline{\yng(1,1)}_3}
\label{twinmattercontent}
\eeq
where now $\overline{P}$ corresponds to the edge between faces 2 and 3, and $\overline{A}$ to the edge between face 3 and its image under the second fixed point. The $SU(N+5)_1$ and $SU(N+5)_3$ gauge invariants are
\beq
\phi_{ab}=\overline{Q}^i_a \overline{Q}^j_b A_{ij}\ \ \ \ , \ \ \ \ \overline{\phi}^{ab}=\overline{P}^a_\alpha \overline{P}^b_\beta \overline{A}^{\alpha\beta}\ ,
\label{gaugeinvSUtwin}
\eeq
where $\alpha,\beta$ are fundamental indices of $SU(N+5)_3$.
They are in the antisymmetric and conjugate antisymmetric representation of $SU(N+1)_2$, respectively. They do not exist for $N=0$, but for $N\geq 1$ the simplest gauge invariant is given by
\beq
\langle \phi_{ab}\overline{\phi}^{ab}\rangle~,
\label{vev_Coulomb_branch}
\eeq
which parametrizes the Coulomb branch in this case. The same remarks as in the previous cases apply. Further, note that this last case allows for a simpler gauge invariant parametrization of the Coulomb branch because it is the only one where the two fixed points (giving rise to $A$ and $\overline{A}$) have the same sign, see \fref{twin_SU5_points}. In the two previous cases the fixed points have opposite signs, and we have to take the loop twice.

\subsubsection*{Double $SU(5)$ Models}

\begin{figure}[h!]
  \centerline{\includegraphics[width=0.47\linewidth]{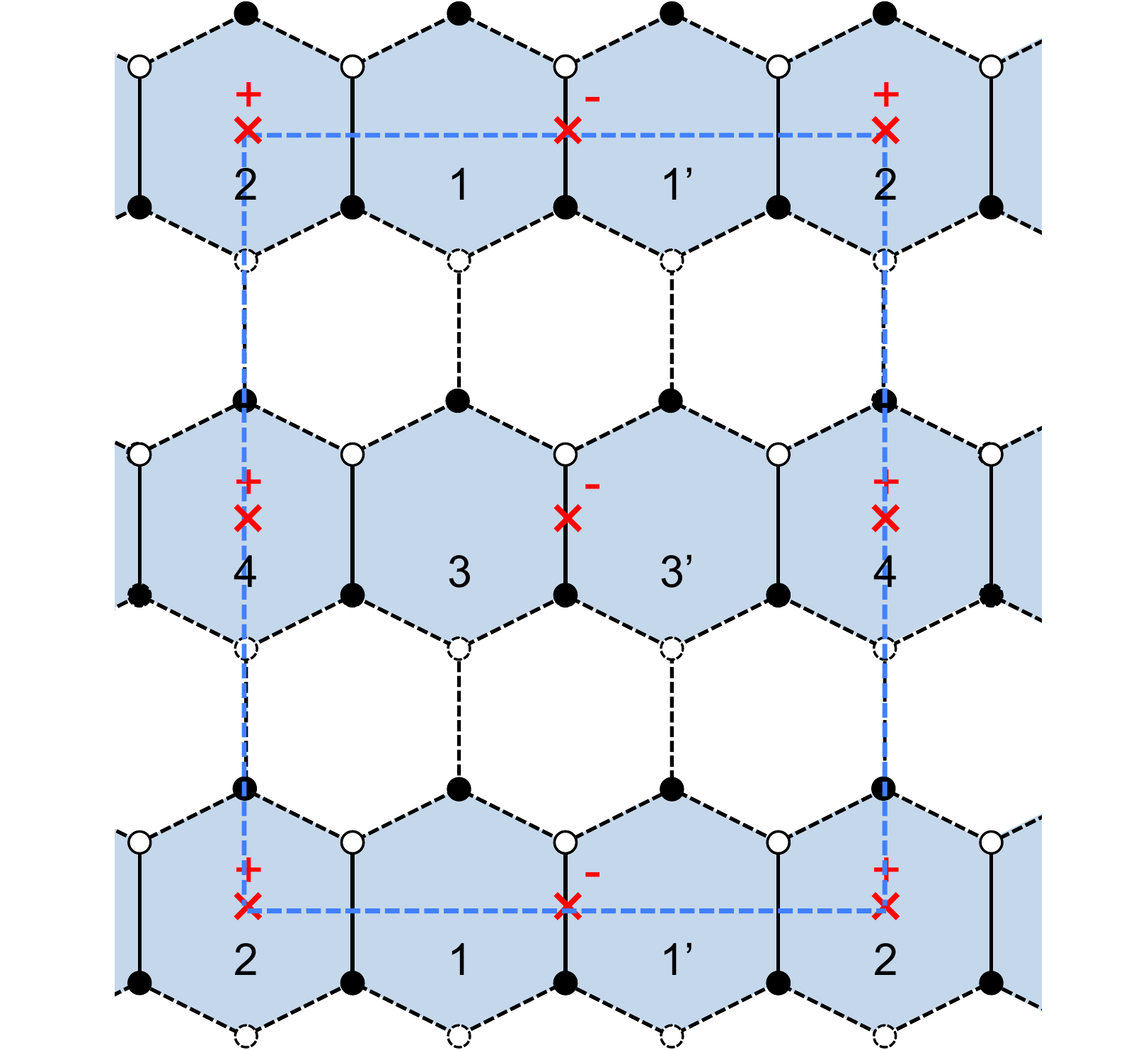}}
    \caption{General structure of a fixed point orientifold realizing a double $SU(5)$ with $SO(1)$ flavor group model.}
\label{double_SU5SO_fixed_points}
\end{figure}

In some cases, the structure of the dimer is such that it could be possible to use all four fixed points to generate a pair of $SU(5)$ models. \fref{double_SU5SO_fixed_points} shows the general structure for a dimer giving rise to two $SU(5)$ models with $SO(1)$ flavor nodes. Other possibilities, for instance two models with $SU(1)$ flavor nodes, an $SU(1)$/$SO(1)$ combination or two twin $SU(5)$ models, are also feasible. The same logic of previous examples applies to each of the two stripes of blue faces, so we conclude that each of these models contain $\mathcal{N}=2$ fractional branes and hence are not stable.

\vskip 10pt

The different cases considered so far illustrate the general strategy that we will apply to most of the other models we will be considering. While the DSB models under consideration are relatively simple, we are considering here their embedding into arbitrarily complicated toric singularities. Therefore, establishing the existence of $\mathcal{N}=2$ fractional branes (which implies the instability of the DSB model) might naively seem an intractable problem since, generically, the majority of the dimer model will be unknown. However, as it occurred in the previous examples, the necessary interplay between the region of the dimer that makes up the DSB model and the orientifold fixed points (or fixed lines, as we will see shortly), implies that we fully know the dimer model along a ``short direction" of the unit cell. This is sufficient to identify an $\mathcal{N}=2$ fractional brane. In even simpler terms, in these cases the DSB models are actually supported on faces of the dimer that define an $\mathcal{N}=2$ fractional brane. We will see that there is only one specific way to circumvent this argument.

\subsection{Fixed Line Orientifolds}

\label{section_SU(5)_fixed_lines}

A second possibility is that dimers admit line reflection. As explained in Appendix \sref{dimers}, we can have orientifolds with either two independent fixed lines or a single diagonal fixed line.

\begin{figure}[h!]
	\centering
	\begin{subfigure}[t]{0.20\textwidth }
		\begin{center} 
			\includegraphics[width=\textwidth]{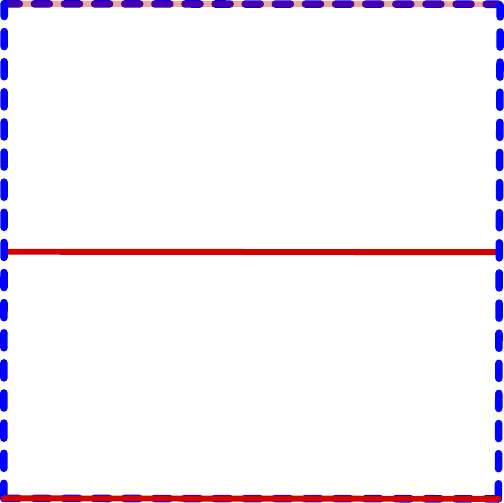}
		\end{center}
	\end{subfigure} \hspace{15mm}
	\begin{subfigure}[t]{0.35\textwidth } 
		\begin{center} 
			\includegraphics[width=\textwidth]{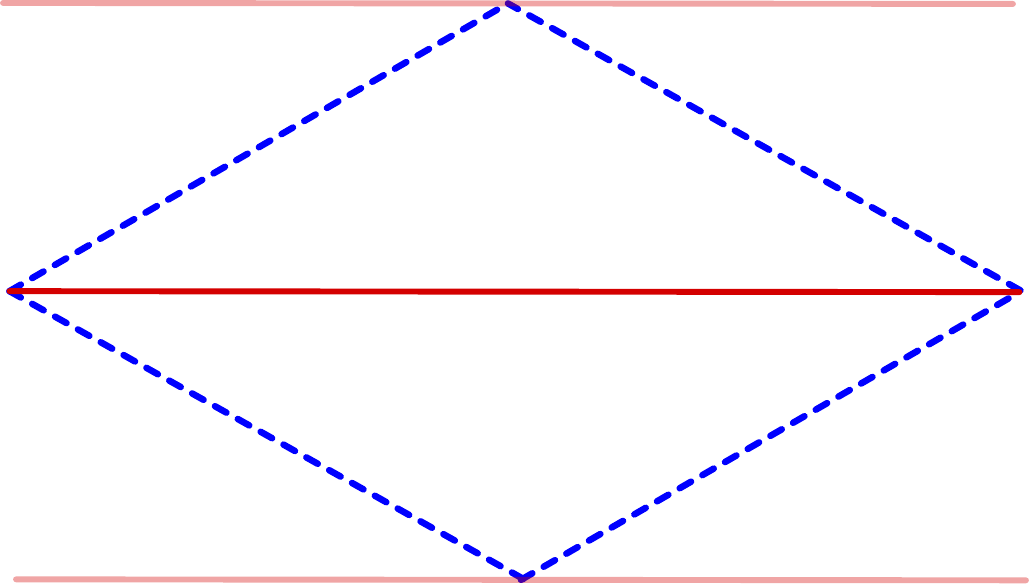}
		\end{center}
	\end{subfigure} 
	\caption{A schematic representation of orientifold fixed lines going through the dimer unit cell: two fixed lines on the left, a single fixed line on the right.}\label{olines} 
\end{figure}

An orientifold with two fixed lines is such that the unit cell of the dimer can be taken to be rectangular, and the dimer is further invariant under a reflection leaving fixed the lines going along one of the boundaries of the unit cell. By the periodicity of the dimer, there must be a second fixed line parallel to the first one, and going through the middle of the unit cell. Vertical and horizontal fixed lines will be considered on the same footing here.

Orientifolds with a single fixed line are such that the unit cell can be taken to have the shape of a rhombus, and the dimer is invariant under reflections about a fixed line which goes along one of the diagonals of the rhombus. The periodicity of the dimer does not imply the presence of other fixed lines in the unit cell. Again, we will not make the distinction between the two diagonals.
Both situations are depicted in \fref{olines}. In the following, we will use the two nomenclatures ``double and single" or ``horizontal/vertical and diagonal fixed lines" interchangeably.

\bigskip

\subsubsection{DSB Models between Two Fixed Lines}

The cases with two fixed lines are basically identical to the orientifolds considered in the previous section, with the exchange of fixed points for fixed lines. We therefore present them succinctly.

\begin{itemize}

\item \underline{$SO$ flavor group}

\fref{SU5SO_lines} shows the local configuration realizing the $SU(5)$ model with $SO(1)$ flavor group, including the signs of the fixed lines. This is achieved by setting $N_1=N_{1'}=5$, $N_2=1$ and vanishing ranks for all other faces. Since the two lines have opposite signs, this configuration is only possible in orientifolds with two independent fixed lines.

\begin{figure}[h!]
	\centerline{\includegraphics[width=0.47\linewidth]{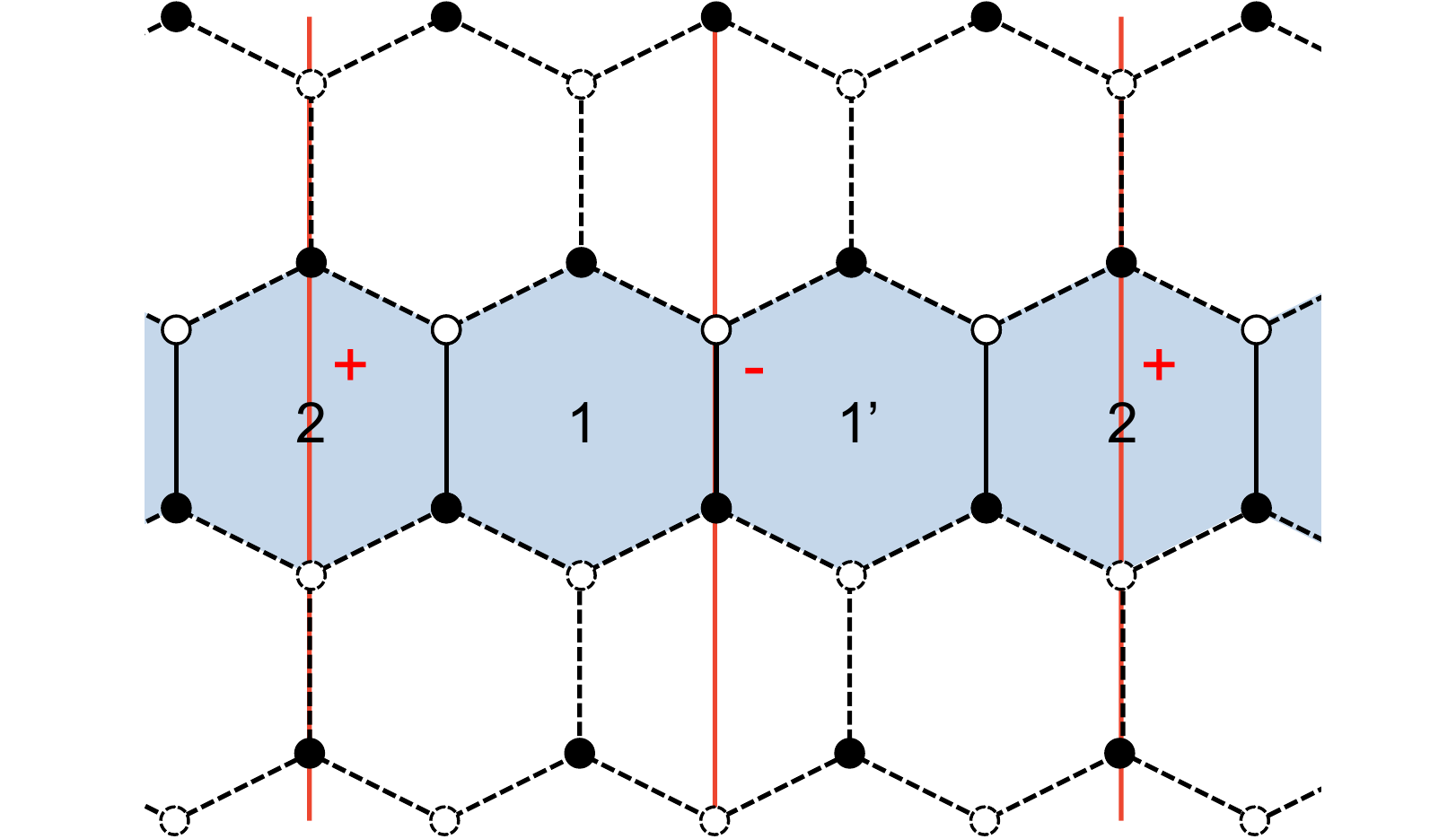}}
	\caption{Two fixed lines orientifold realizing the $SU(5)$ model with $SO(1)$ flavor group.}
	\label{SU5SO_lines}
\end{figure}

\newpage

\item \underline{$SU$ flavor group with symmetric}

\fref{SU5SU_lines} shows the local configuration realizing the $SU(5)$ model with $SU(1)$ flavor group and a symmetric. This corresponds to $N_1=N_{1'}=5$, $N_2=N_{2'}=1$ and vanishing ranks for all other faces.
Since the two lines have opposite signs, this configuration is only possible in orientifolds with two independent fixed lines.

\begin{figure}[h!]
  \centerline{\includegraphics[width=0.47\linewidth]{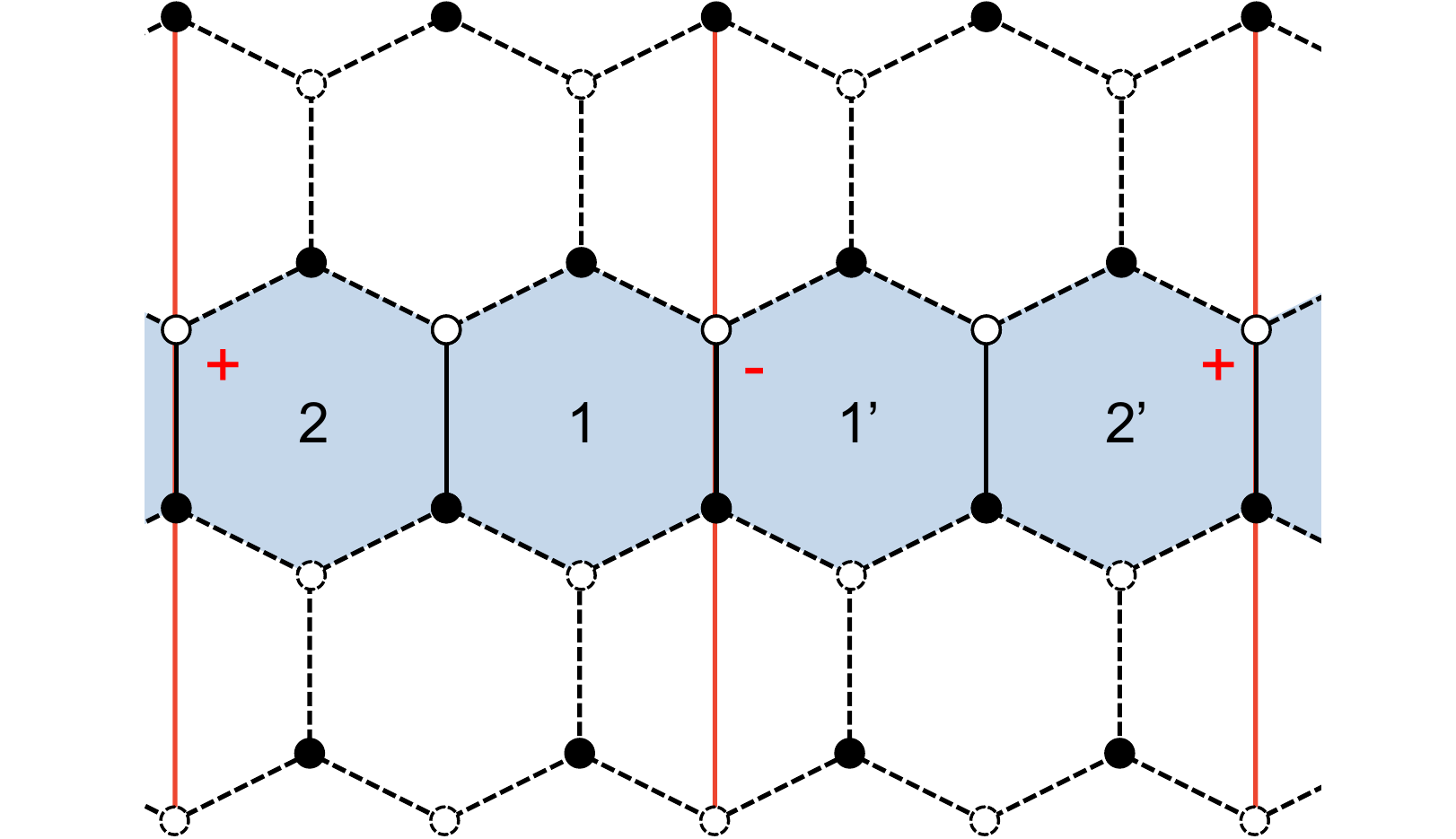}}
    \caption{Two fixed lines orientifold realizing the $SU(5)$ model with $SU(1)$ flavor group.}
\label{SU5SU_lines}
\end{figure}

\item \underline{$SU$ flavor group with twin $SU(5)$}

\fref{twin_SU5_lines} shows the local configuration realizing the $SU(5)$ model with $SU(1)$ flavor group and a twin $SU(5)$ model. This corresponds to $N_1=N_{1'}=5$, $N_2=N_{2'}=1$, $N_{3}=N_{3'}=5$ and vanishing ranks for all other faces.
In this case the two lines have the same sign, hence it is possible to find this configuration both in orientifolds with two independent fixed lines or with a single diagonal fixed line. Note that in the latter case, we have to consider the situation in which the strip goes from one line to a second one, in a contiguous unit cell. 

\begin{figure}[h!]
	\centerline{\includegraphics[width=0.63\linewidth]{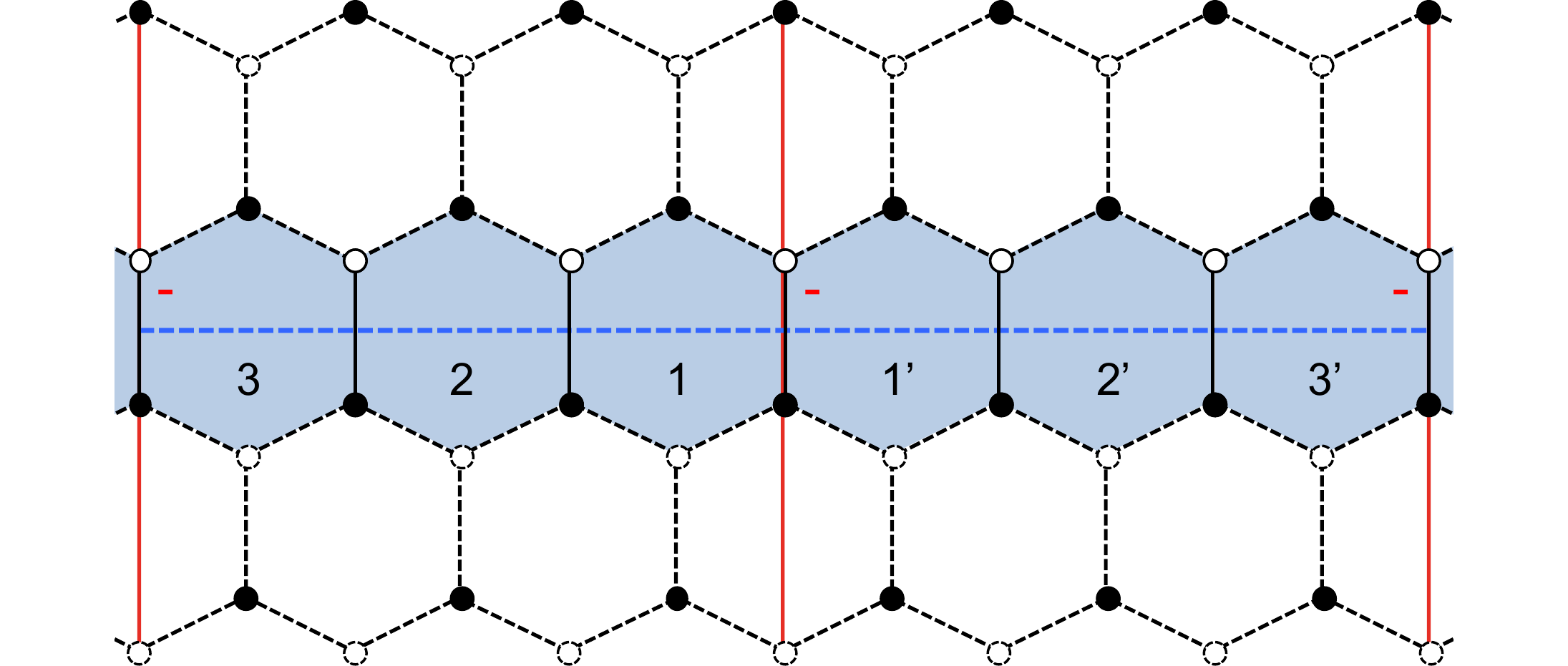}}
	\caption{Two fixed lines orientifold realizing the twin $SU(5)$ model.}
	\label{twin_SU5_lines}
\end{figure}

\end{itemize}

Using the same arguments as for the fixed point orientifolds in \sref{section_SU(5)_fixed_points}, we conclude that in all these cases the models are supported on a stripe of faces of the dimer that define an $\mathcal{N}=2$ fractional brane.

\subsubsection*{Multiple $SU(5)$ Models}

\begin{figure}[h!]
  \centerline{\includegraphics[width=0.47\linewidth]{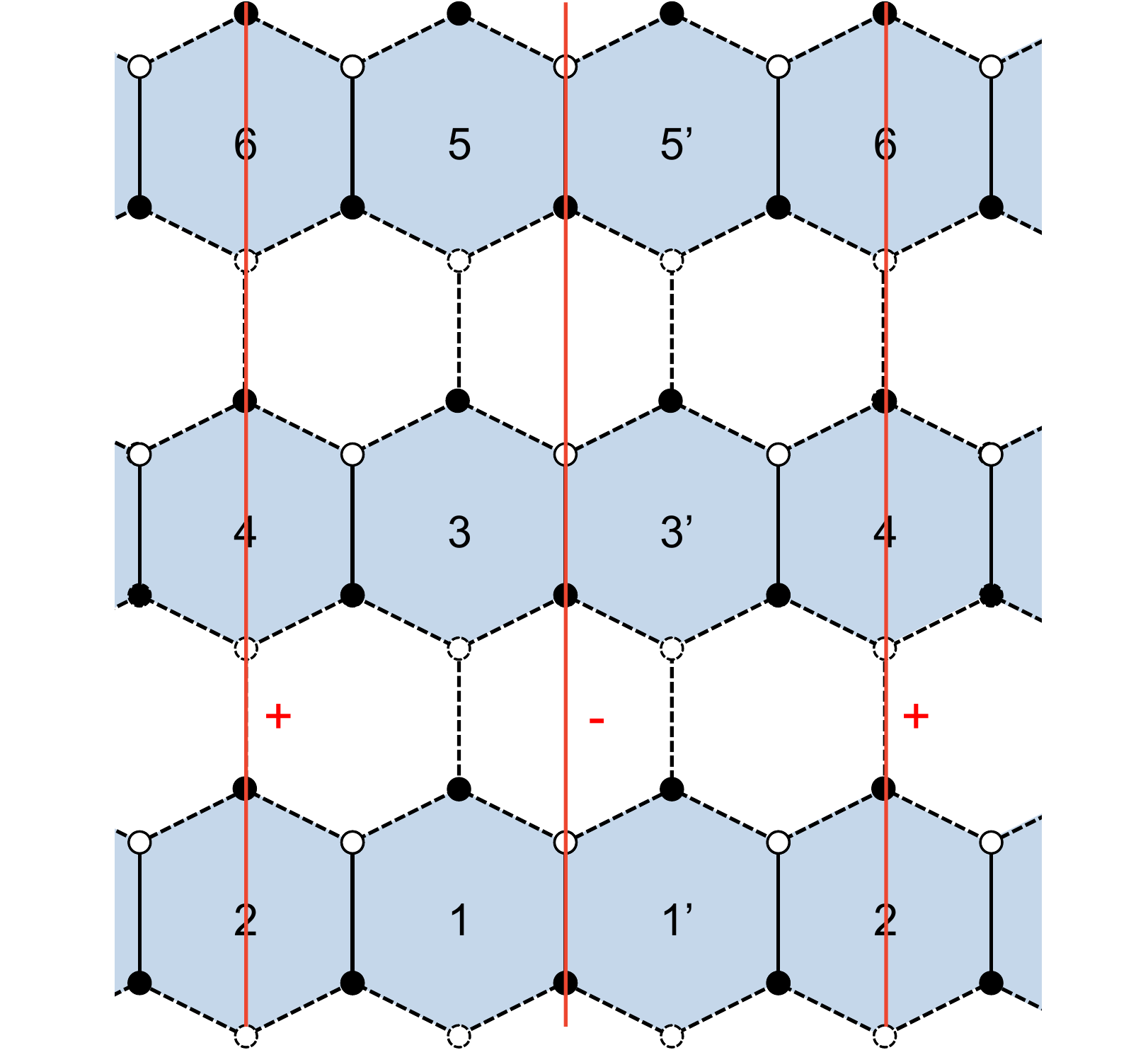}}
    \caption{An example of the general structure of a portion of a dimer with two fixed lines giving rise to multiple $SU(5)$ models.}
\label{multiple_SU5_fixed_lines}
\end{figure}

We previously saw that fixed point orientifolds can give rise to double $SU(5)$ models. Similarly, orientifolds with fixed lines can produce multiple $SU(5)$ models, as shown in \fref{multiple_SU5_fixed_lines}. In this case, the number of models is not restricted to two. It is important to note that, unlike in the example shown in the Figure, it is possible for different stripes to use the two fixed lines in different ways, for instance simultaneously leading to models with both $SO(1)$ and $SU(1)$ flavor groups, when the two lines have opposite signs. Once again, our general discussion applies to each individual stripe of blue faces, so we conclude that $\mathcal{N}=2$ fractional branes exist for each individual stripe and hence the models are not stable.

\subsubsection{DSB Models on a Single Fixed Line: the Twin $SU(5)$}

There is one additional way in which an $SU(5)$ model could be engineered. This is when both the projection needed for the antisymmetric of $SU(5)$ and the one for canceling the anomaly due to the antifundamental, are provided by the same fixed line. 
This could be realized both in orientifolds with a diagonal fixed line, and in orientifolds with two fixed lines. What is important is that only one line is needed to define the relevant cluster of faces.

Importantly, since the orientifold line cannot change sign along the dimer, this possibility is effective only when the two projections have the same sign. Then the only case that fits the bill is the twin $SU(5)$ model, as the one in \fref{twin_SU5_lines}. 

\begin{figure}[h!]
	\centerline{\includegraphics[width=0.27\linewidth]{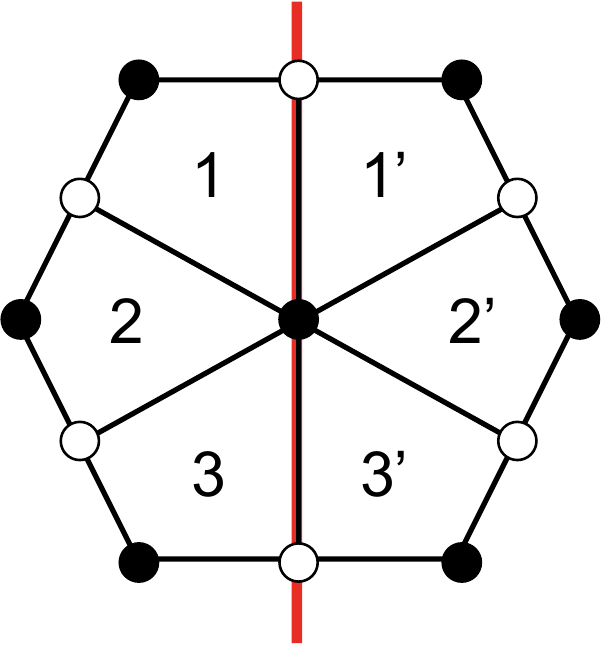}}
	\caption{The hexagonal cluster with six faces on an orientifold line. All faces are here depicted with four edges, but some of them could have more.}
	\label{hexagon}
\end{figure}

Basically, the chain of gauge groups represented by faces 1, 2 and 3 has to bend and end on the same line. There are now two possibilities. Either all the black nodes at the bottom of the edges between faces 1, 2 and 3 are one and the same, or the chain 1-2-3 and their images enclose some (unoccupied) faces of the dimer. The latter case is inconsistent from the dimer point of view, as shown in Appendix \ref{dimers}: such a chain cannot be a fractional brane in the parent theory. We are thus left with the former case, which in the dimer corresponds to a hexagonal cluster of faces around a node, as depicted in \fref{hexagon}.

Interestingly, such a collection of faces surrounding a node corresponds to a deformation fractional brane in the classification of \cite{Franco:2005zu}. It is reassuring that unlike in the cases with fixed points, deformation branes are compatible with line orientifolds \cite{Retolaza:2016alb}.  

The analysis of this case is similar to what we carried out for the twin $SU(5)$ model previously, leading to a gauge group
\beq
SU(N+5)_1 \times SU(N+1)_2 \times SU(N+5)_3 \ .
\eeq
The difference is that now the node at the center of the hexagonal cluster corresponds to a sextic superpotential term. Using the same notation as in \eref{twinmattercontent}, we have 
\begin{equation}\label{superpotentialtwin}
W = \tr A \overline{Q}\overline{P}\overline{A}\overline{P}^t \overline{Q}^t = \tr \phi \overline{\phi}\ .
\end{equation}
For $N=0$, the superpotential vanishes and we are left with two $SU(5)$ models sharing an $SU(1)$ flavor node, in which both surviving $SU(5)$ factors break supersymmetry dynamically into a stable vacuum. Unlike the other realizations of the twin $SU(5)$ model, in the present one there is no indication that the dimer must contain an $\mathcal{N}=2$ fractional brane.

\bigskip

Combining the analysis in \sref{section_SU(5)_fixed_points} and \sref{section_SU(5)_fixed_lines}, we conclude that engineering a single $SU(5)$ DSB model without instabilities at an orientifold of a toric singularity is impossible. Conversely, our analysis implies that engineering a minimal $SU(5)$ model requires non-isolated singularities with curves of $\mathbb{C}^2/\mathbb{Z}_n$ singularities passing through the origin, which in turn result in the instability. This means that, as explained in \sref{section_N=2_dimers}, the toric diagram must contain internal points on its boundary edges.  On the other hand, our analysis shows that an instance of a DSB model, the twin $SU(5)$ model, actually exists which is compatible with an orientifold projection with fixed line(s). We should now understand whether such sub-dimer can actually be embedded into a consistent dimer and, if so, whether such dimer can be free of $\mathcal{N}=2$ fractional branes. We investigate these questions in \sref{Octagon}.

\section{3-2 Models}
\label{32}

Let us now turn to the 3-2 model, another prominent example of DSB that was recovered within brane setups at orientifold singularities in \cite{Argurio:2019eqb}. The model has gauge group $SU(3) \times SU(2)$. Its matter content is reminiscent of one SM generation
\begin{eqnarray}
Q=(\tiny  {\yng(1)}_3 , \tiny \overline{\yng(1)}_2)~~,~~\wb U=\tiny \overline{\yng(1)}_3~~,~~\wb D=\tiny \overline{\yng(1)}_3 ~~,~~L=\tiny{\yng(1)}_2~,
\end{eqnarray}
where the subindices indicate the corresponding gauge group in an obvious way. In addition, the theory has the following superpotential
\beq
W=\wb D Q L ~.
\label{W_32}
\eeq

In principle, the above field content ($SU$ gauge groups, (bi)fundamental matter, together with a cubic superpotential) does not seem to require an orientifold projection. As it will become clear in the following, such a projection is nevertheless necessary in order to allow for a fractional brane (i.e.~an anomaly free configuration) with the desired ranks for the gauge groups.

\subsection{General Features} 

Let us think more carefully about the basic features of the D-brane realization of this model. In this subsection we enumerate all different ways to recover the 3-2 model from fractional branes at an orientifold singularity. The structure of these models is more intricate than that of the $SU(5)$ model, so it is convenient to draw the corresponding quivers.

\begin{figure}[h!]
	\centerline{\includegraphics[height=5.75cm]{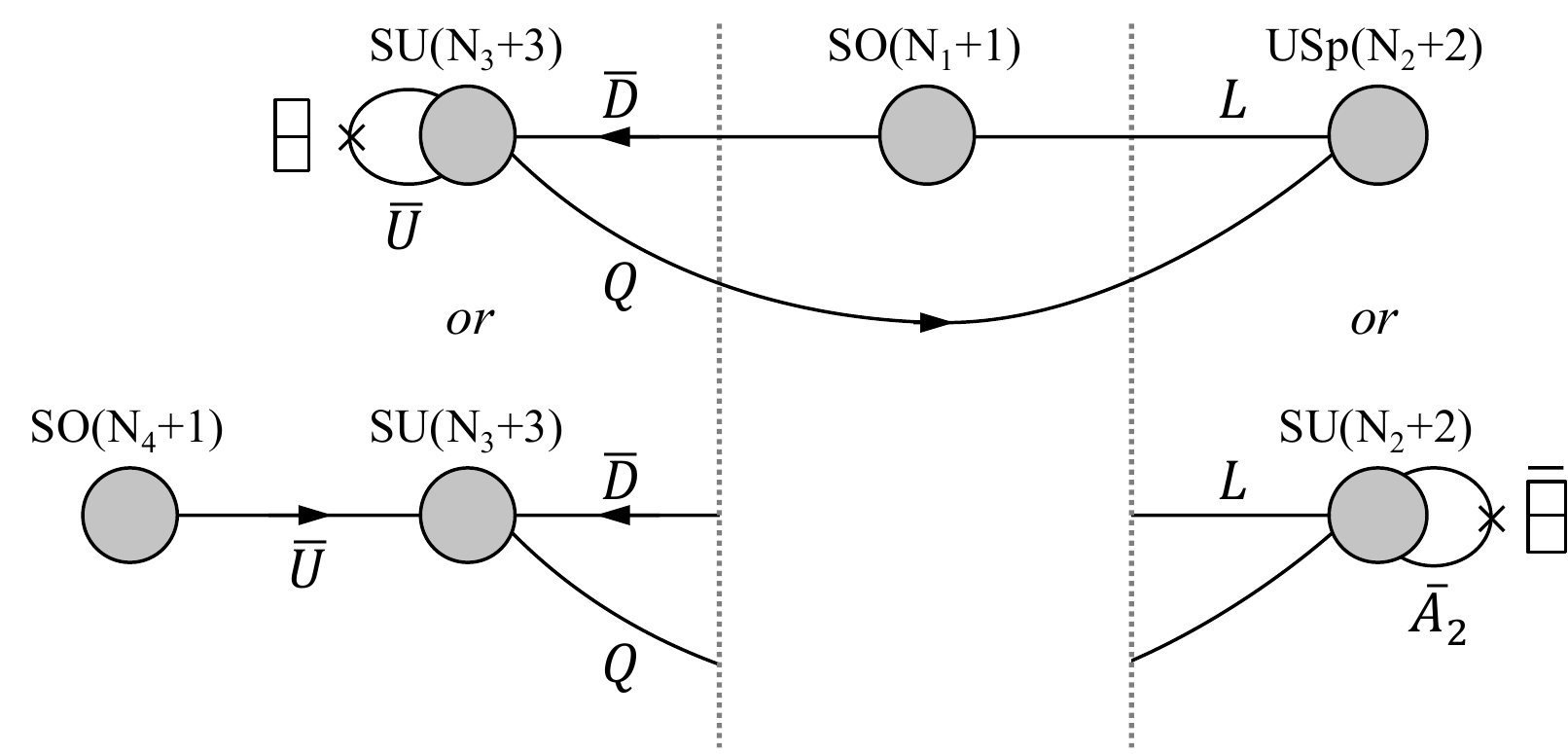}}
	\caption{Four quivers giving rise to the 3-2 model when all $N_i =0$. All these models use three orientifold fixed loci.} 
	\label{quiver_32_1}
\end{figure}

The candidate models are presented in \fref{quiver_32_1}. In the figure, we have kept the ranks of the gauge group general by introducing $N_i$, $i=1,\ldots, 4$. These additional integers account for more general configurations of D-branes at the singularity, e.g. the addition of regular or fractional D3-branes, and we posit that anomaly cancellation must hold even in those cases. The 3-2 model arises when all $N_i$ and the ranks of additional gauge groups, which depend on the specific singularity and are not shown in these quivers, vanish.

For similar reasons as in the case of the $SU(5)$ model, we need at least an additional gauge group factor, which we will call node 1, to serve as a flavor group providing the $\wb D$ and $L$ fields. Both $\wb D$ and $L$ should be connected to the same node for the superpotential \eref{W_32} to be possible. In dimer terminology, we identify the smallest building block of a 3-2 model as three faces connected by a trivalent vertex. In this sense 3-2 model realizations are necessarily more involved than $SU(5)$ model realizations, since the latter only required a building block of two faces.

The quivers in \fref{quiver_32_1} should be interpreted as follows. For each of the two endpoints of the quiver, we have presented two possibilities. The two options on the left correspond to realizing $\wb U$ as an antisymmetric of node $SU(3)$ or via a fourth gauge group acting as a flavor node. The two options on the right correspond to the fact that node 2 can be either $USp(2)$ or $SU(2)$.  All possible combinations of these endpoints realize the desired 3-2 model, therefore \fref{quiver_32_1} accounts for four models. 

In principle, the flavor nodes 1 and 4 in \fref{quiver_32_1} could be $SU$ or $SO$. However, if these nodes were of $SU$ type, their ACC in the case of general ranks would require additional nodes, that come to life when regular D-branes are added. Generically, these gauge groups will give rise to new matter fields charged under the nodes of the original quiver. Such fields would contribute to and potentially help in the cancellation of anomalies. However, for $N$ regular D3-branes, it is easy to show that for neither node the anomaly would cancel, as there would still be an imbalance of one or three units for nodes 1 and 4, respectively. In order to cancel the anomalies there are then only two options. The first is to introduce an orientifold projection. It turns out that setting both nodes to be $SO$ is the simplest such option, and without loss of generality we will stick to it in the following. The second option is to compensate the anomaly by a mirror construction. We defer the treatment of the latter possibility to the last subsection.

It is worth noting that in two of the four models described by \fref{quiver_32_1}, those for which the second gauge group is $SU(N_2 +2)$, we have also introduced an antisymmetric tensor $\bar{A}_2$. This field is necessary for satisfying the ACC for the more general ranks that arise when regular D3-branes are added (see Appendix \ref{appendix_ACC_32}). It becomes a singlet when $N_2 =0$, so it decouples and does not affect the IR physics.

A final option is to get the two antifundamentals of the $SU(3)$, $\wb U$ and $\wb D$ from the same flavor $SO(1)$ group. However, in order to realize the 3-2 model, the structure of the dimer model should be such that a $\wb U Q L $ term is not present in the superpotential. This possibility is then obtained by simply identifying nodes 1 and 4 in \fref{quiver_32_1}.

We thus reach the conclusion that we need no less than three orientifold projections to realize a 3-2 model: one for the $SO(1)$ flavor group (thus with a $+$ sign), one for node 2 which is either $USp(2)$ or $SU(2)$ with an antisymmetric (in both cases, with a $-$ sign), and one for node 3, either with an antisymmetric ($-$ sign) or with the $SO(1)$ flavor node 4 ($+$ sign). Of course some of these projections can be given by the same object, in the case of an orientifold line, provided they require the same sign.\footnote{It is worth noting that in all the realizations of the 3-2 model found in \cite{Argurio:2019eqb}, node 3 has an antisymmetric, node 1 is of $SO$ type, while node 2 is $USp(2)$ in the $\mathbb{Z}_6'$ orbifold and in $PdP_4$, and $SU(2)$ with an antisymmetric in $PdP_{3c}$, $PdP_{4b}$ and the $\mathbb{Z}_3 \times \mathbb{Z}_3$ orbifold.}

All quivers described by Figure \ref{quiver_32_1} are viable as stand-alone gauge theories. However, as for the $SU(5)$ model, we need to verify whether the theories remain anomaly free upon the addition of regular and/or fractional D3-branes. It turns out that the $SO(N_1 +1)\times SU(N_2 +2) \times SU(N_3 +3) \times SO(N_4 +1)$ model does not pass this test, 
as shown in Appendix \ref{appendix_ACC_32}.

Below we investigate the realization of these models in terms of fixed point and fixed line orientifolds.

\subsection{Fixed Point Orientifolds}

\label{section_32_fixed_points}

Interestingly, for the purpose of establishing the existence of an $\mathcal{N}=2$ fractional brane, and hence the instability of the supersymmetry breaking vacuum, it is sufficient to focus on a very small part of all these theories. In particular, all of them contain one of the following two subsectors: 

\begin{itemize}
\item $SO(N_1 +1) \times USp(N_2 +2)$.
\item $SO(N_1 +1) \times SU(N_2 +2)$ with the tensor $\wb A_2$. 
\end{itemize}
\noindent Knowledge of the dimer around gauge groups 1 and 2 will be enough for our purposes. Let us consider the general structure of the dimers associated to these two possibilities.

\begin{itemize}

\item \underline{$SO(N_1 +1) \times USp(N_2 +2) \subset$ 3-2 model}

\fref{dimer_32_points_SOUSp} shows the general structure of the relevant part of the dimer model. The edge between faces 1 and 2 represents the $L$ field. Clearly, faces 1 and 2 define a stripe that winds around the unit cell of the parent dimer, giving rise to a gauge invariant that is not in the superpotential. Therefore, they correspond to an $\mathcal{N}=2$ fractional brane. 

\begin{figure}[h!]
  \centerline{\includegraphics[width=0.36\linewidth]{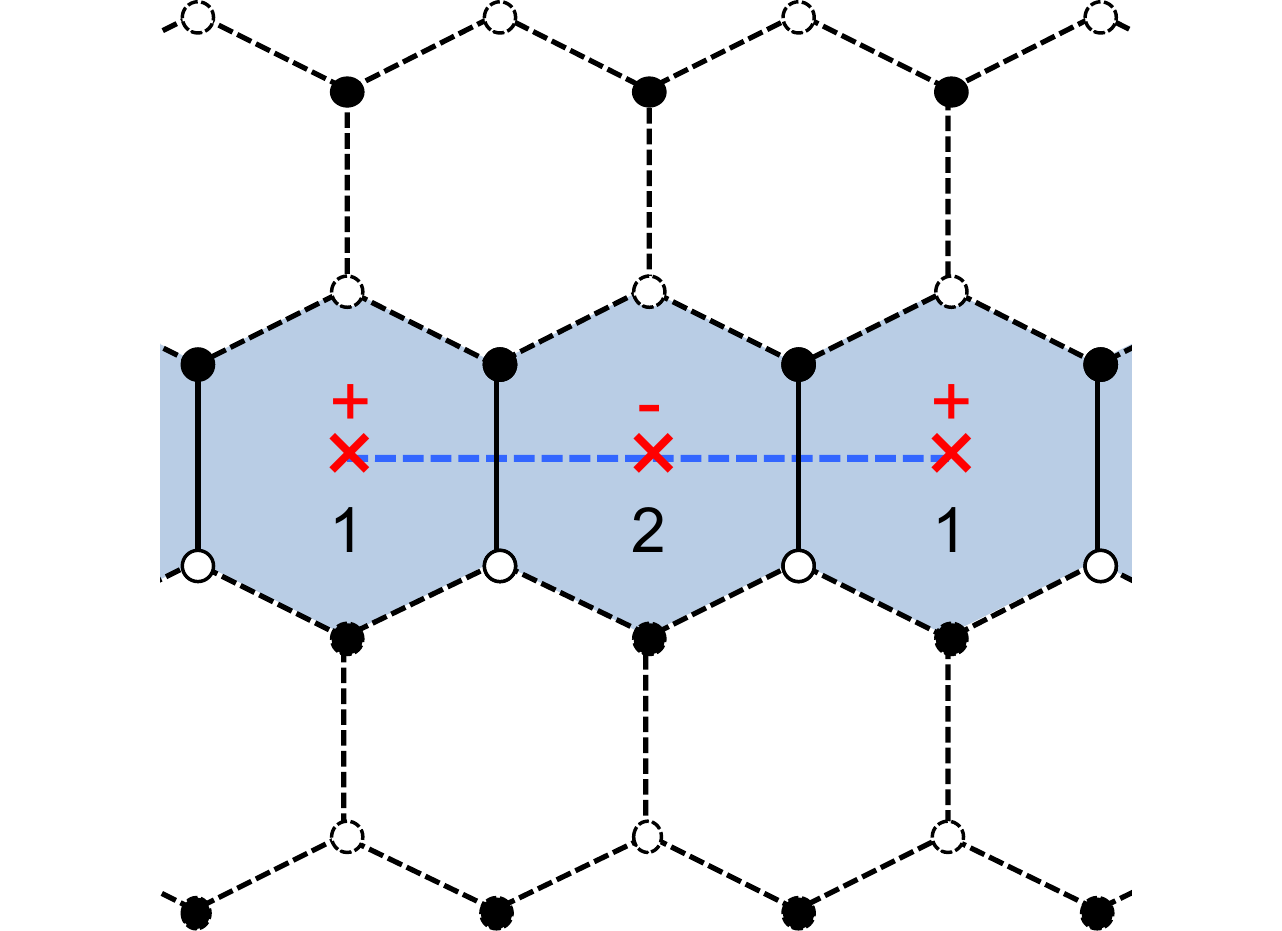}}
    \caption{A piece of the dimer for a fixed point orientifold realizing the 3-2 model with an $SO(N_1 +1) \times USp(N_2 +2)$ subsector.}\label{dimer_32_points_SOUSp}
\end{figure}

\item \underline{$SO(N_1 +1) \times SU(N_2 +2)$ with $\wb A_2$ $\subset$ 3-2 model}

\fref{dimer_32_points_SOA} shows the part of the dimer that we are interested in. The edge between faces 1 and 2 corresponds to $L$, while the one between face 2 and its image gives rise to $\wb A_2$. Once again, we see that faces 1, 2 and 2' define an $\mathcal{N}=2$ fractional brane in the parent dimer. It is interesting to note that this picture is identical to \fref{SU5SO} for the $SU(5)$ model. 

\begin{figure}[h!]
  \centerline{\includegraphics[width=0.47\linewidth]{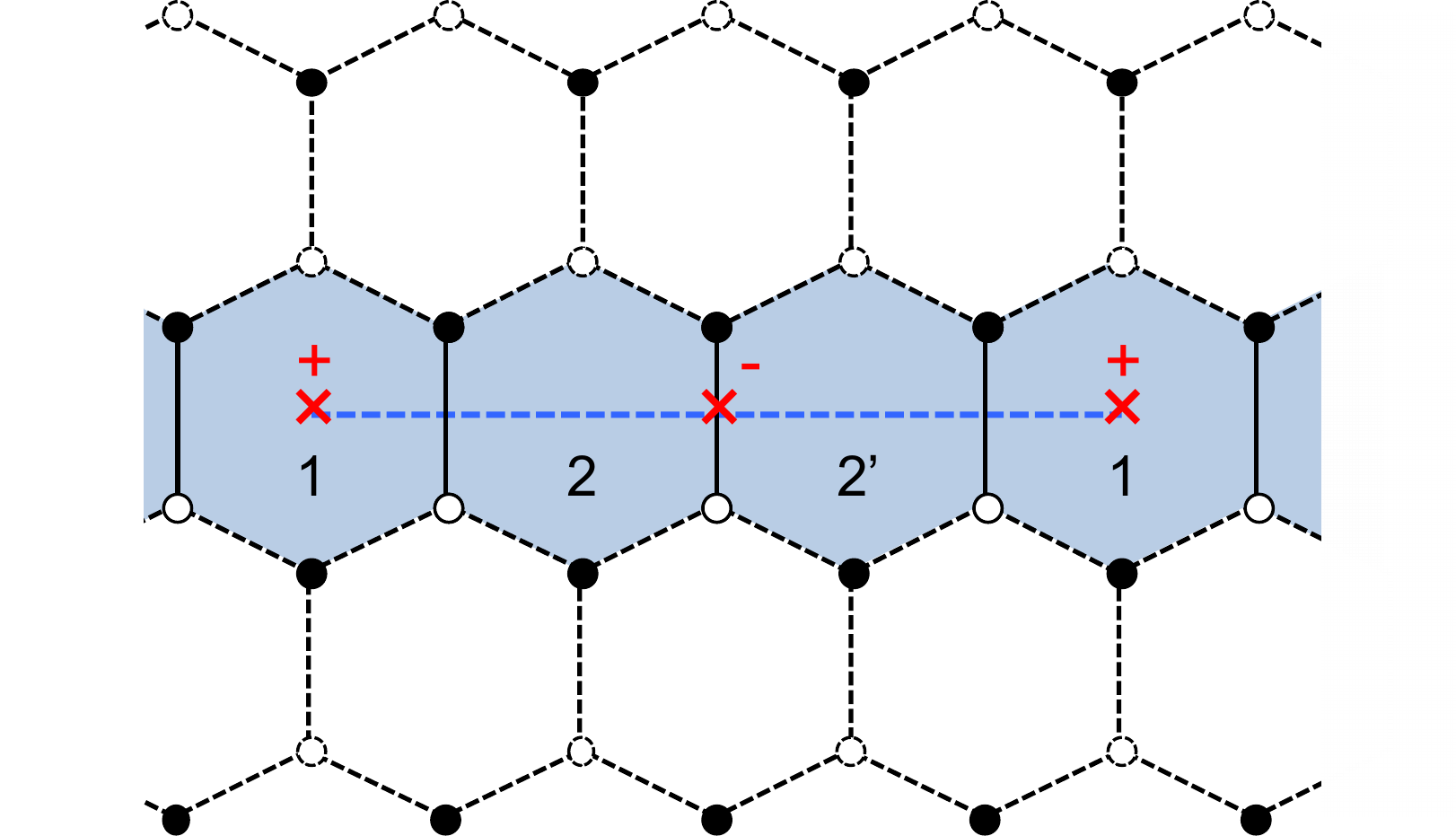}}
    \caption{A piece of the dimer for a fixed point orientifold realizing the 3-2 model with an $SO(N_1 +1) \times SU(N_2 +2)$ with $\wb A_2$ subsector.}
    \label{dimer_32_points_SOA}
\end{figure}

\end{itemize}

From the previous discussion, we conclude that all realizations of the 3-2 model at fixed point orientifolds suffer from an $\mathcal{N}=2$ fractional brane instability.

\subsubsection*{Models with more than one type of $\mathcal{N}=2$ fractional branes}

\begin{figure}[h!]
  \centerline{\includegraphics[width=0.42\linewidth]{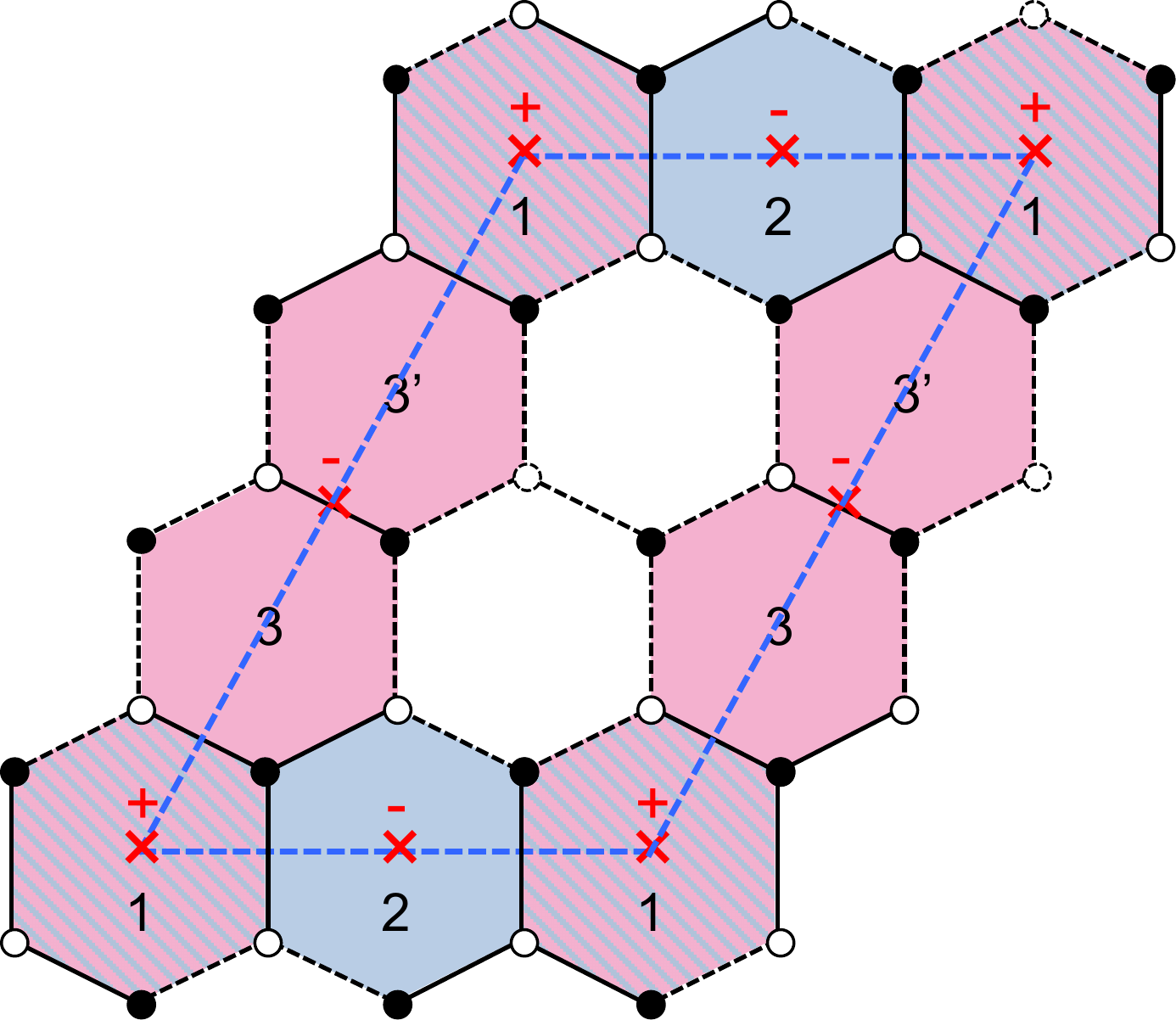}}
    \caption{General structure of the dimer model for one of the models in \fref{quiver_32_1}. This model contains two different $\mathcal{N}=2$ fractional branes. They are shown in blue and pink, with the striped face belonging to both of them.
}
\label{dimer_32_2branes}
\end{figure}

Before moving on, let us consider the models in Figures \ref{dimer_32_points_SOUSp} and \ref{dimer_32_points_SOA} in further detail. As we have already mentioned, in all these cases the portion of the dimer realizing the 3-2 model involves three fixed points. For concreteness, let us focus on the case in which $\wb U$ is an antisymmetric of node 3 and node 2 if of $USp$ type. All other combinations are analogous and lead to the same conclusions. \fref{dimer_32_2branes} shows the general structure of the dimer model. Interestingly, in this case we can identify yet another $\mathcal{N}=2$ fractional brane, in addition to the one covered by our previous analysis. This new fractional brane corresponds to faces 1, 3 and 3' in the parent dimer and is shown in pink in \fref{dimer_32_2branes}. We conclude that when sub-dimers as in Figures \ref{dimer_32_points_SOUSp} and \ref{dimer_32_points_SOA} are embedded in a complete dimer model,  the corresponding toric singularity has at least two different types of $\mathcal{N}=2$ fractional branes. Explicit models illustrating this phenomenon were constructed in \cite{Argurio:2019eqb}.

Another interesting fact we would like to notice has to do with the intertwining between $SU(5)$ and 3-2 models realizations. \fref{dimer_32_2branes} shows that in any such configuration realizing a 3-2 model, an $SU(5)$ model can also be realized, by simply turning off the rank of node 2, while pumping up the rank of node 3 to $SU(5)$. Even more, 3-2 model realizations like the one of \fref{dimer_32_points_SOA} allow for two alternative $SU(5)$ model realizations, the other one being by turning off node 3 and setting node 2 to $SU(5)$, as already noticed when commenting the figure. Multiple explicit examples of this connection can be found in \cite{Argurio:2019eqb}. The only realization of a 3-2 model that does not lead directly to a realization of the $SU(5)$ model would be one with $USp(2)$ at node 2 and a node 4 to compensate the anomaly of node 3. Unfortunately, no examples of this exist in the literature, and it is beyond our scope to find one here, as we have in any case shown that it would be afflicted by an $\mathcal{N}=2$ fractional brane instability.

\subsubsection*{Double 3-2 Models}

It is natural to ask whether fixed point orientifolds can lead to a pair of 3-2 models. In this case, each of the models should use two of the four fixed points. However, all the models of \fref{quiver_32_1} need three different projections, and thus three different fixed points. One could still think about the case where nodes 1 and 4 are identified, where only two identifications are actually required. However in order for node 3 to have two different connections with node 1, the faces corresponding to this 3-2 model realization end up being spread across all the unit cell, so that again two such models cannot coexist.\footnote{It would be interesting to investigate whether such model can actually be engineered in terms of dimers. Again, since we have already proven that all realizations of the 3-2 models at fixed point orientifolds are unstable, we do not pursue this challenging question any further.}

\subsection{Fixed Line Orientifolds}

\label{section_32_fixed_lines}

We now consider the realization of the 3-2 models in orientifolds with fixed lines.

The analysis in the case in which the 3-2 model uses two different orientifold fixed lines is identical to the one for fixed points. In particular, it is sufficient to focus on faces 1 and 2. We simply need to replace fixed points by fixed lines in the previous discussion.

\begin{itemize}

\item \underline{$SO(N_1 +1) \times USp(N_2 +2) \subset$ 3-2 model}

\fref{dimer_32_lines_SOUSp} shows the relevant part of the dimer. We immediately identify an $\mathcal{N}=2$ fractional brane in the parent dimer consisting of faces 1 and 2.

\begin{figure}[h!]
  \centerline{\includegraphics[width=0.36\linewidth]{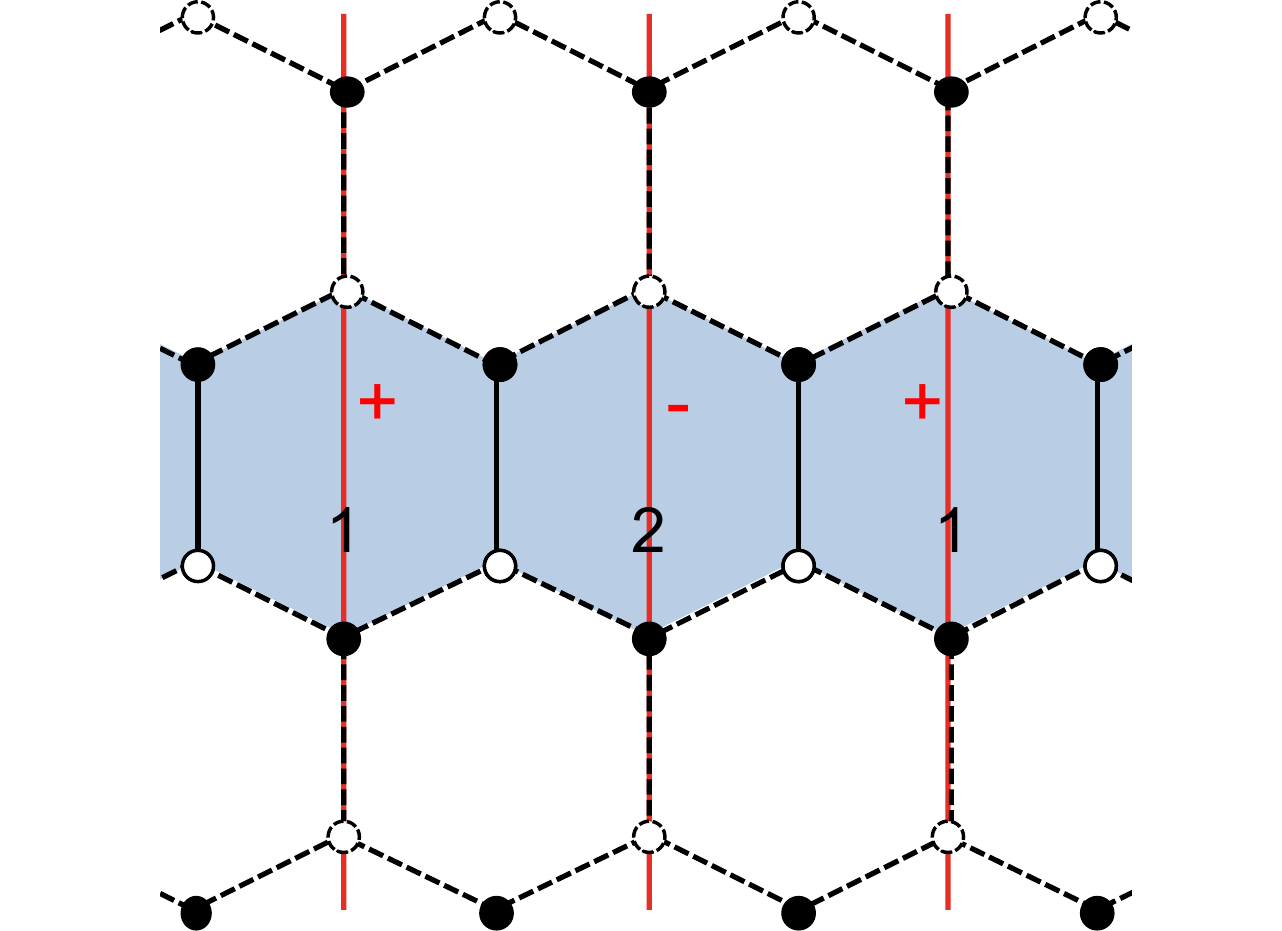}}
    \caption{A piece of the dimer for an orientifold with two fixed lines realizing the 3-2 model with an $SO(N_1 +1) \times USp(N_2 +2)$ subsector.}
\label{dimer_32_lines_SOUSp}
\end{figure}

\item \underline{$SO(N_1 +1) \times SU(N_2 +2)$ with $\wb A_2$ $\subset$ 3-2 model}

\fref{dimer_32_lines_SOA} shows the part of the dimer that we focus on. Faces 1, 2 and 2' form an $\mathcal{N}=2$ fractional brane in the parent dimer.

\begin{figure}[h!]
  \centerline{\includegraphics[width=0.47\linewidth]{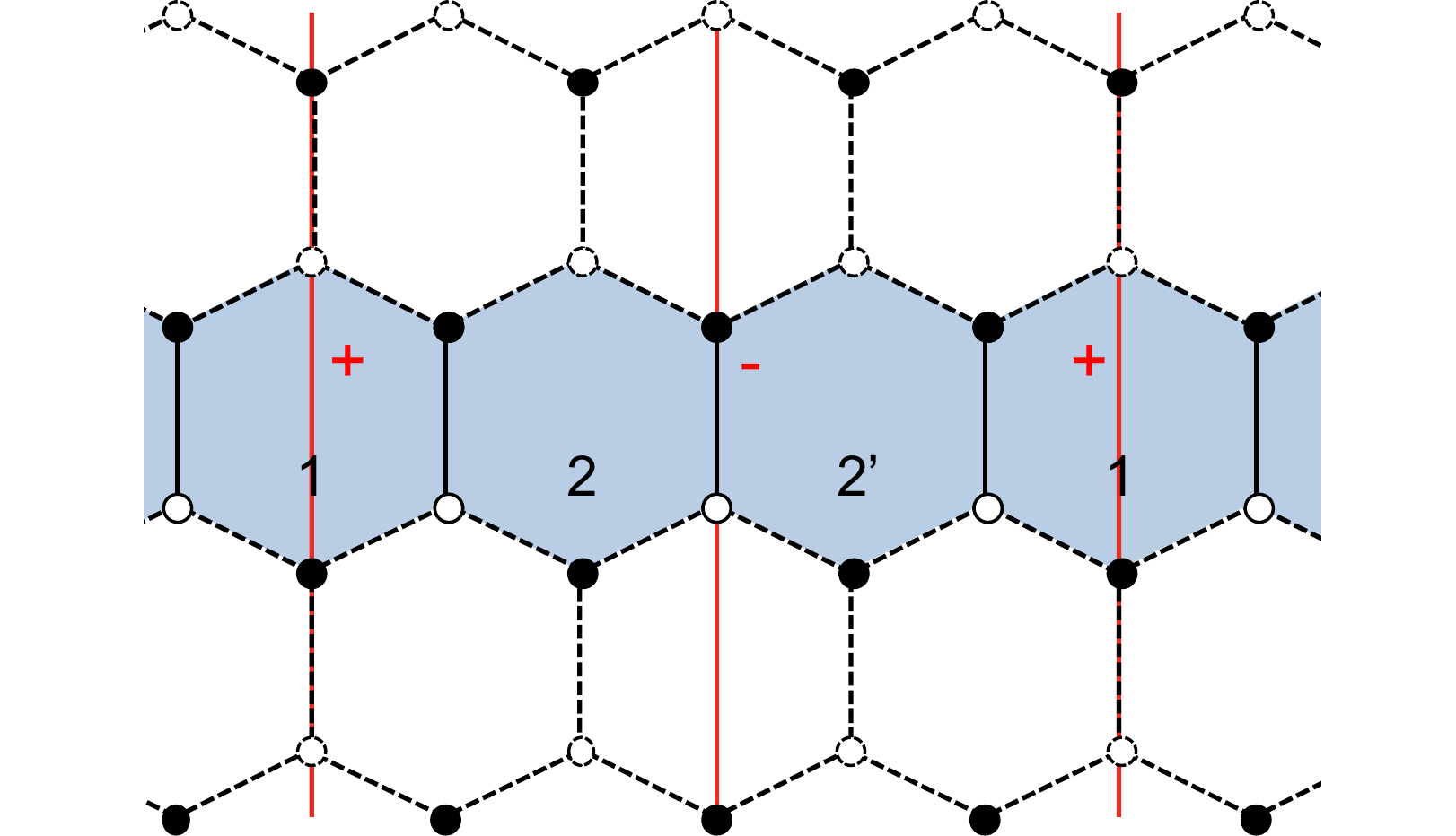}}
    \caption{A piece of the dimer for an orientifold with two fixed lines realizing the 3-2 model with an $SO(N_1 +1) \times SU(N_2 +2)$ with $\wb A_2$ subsector.}
\label{dimer_32_lines_SOA}
\end{figure}

\end{itemize}

\subsubsection*{Multiple 3-2 Models}

Orientifolds with fixed lines can in principle give rise to multiple 3-2 models, stacking them as we did in \fref{multiple_SU5_fixed_lines} for $SU(5)$. In this case, the projection needed for node 3 can be provided either by the line with a $-$ sign, in case of an antisymmetric, or by the line with a $+$ sign, in case of a flavor node 4.
Our previous arguments show that each of these models contain (at least) an $\mathcal{N}=2$ fractional brane and are hence unstable.

\subsubsection*{SU(5) - 3-2 Mixed Models}

At this point it is interesting to point out that our arguments for multiple models, in the case of fixed lines, indicate that we can also have models that realize a combination of $SU(5)$ and 3-2 models. Once again, our arguments from \sref{su5} and this section show that each DSB sector would be independently unstable.

\subsection{Twin 3-2 models?}

We are now left to investigate the possibility that the anomalies of the 3-2 model are cancelled in a twin realization, along the lines of what was done for the $SU(5)$ model in Figures  \ref{twin_SU5_points} and \ref{twin_SU5_lines}. Further, we would like to know if there is a realization similar to the one of \fref{hexagon}, i.e.~on a {\it single} fixed line, which would not automatically imply the presence of $\mathcal{N}=2$ fractional branes.

As already alluded to, we can cancel the anomalies of a node 1 of $SU$ nature, and/or node 4, if in the configuration there is a twin copy of the 3-2 model sharing the $SU(1)$ node. Note that in compensating the anomaly with a twin, it is important that the two models are decoupled. If we were to use the same mechanism to compensate the anomaly of node 2, the non-zero coupling of node 2 itself would couple the twins and alter the low-energy physics of the models (typically destroying the stable supersymmetry breaking dynamics). Hence whatever we do, node 2 will always require a projection. As a consequence, if such twin model is realized in a way that it extends between two different fixed points or fixed lines, by the same arguments used around Figures  \ref{twin_SU5_points} and \ref{twin_SU5_lines}, there will be $\mathcal{N}=2$ fractional branes that render the DSB model eventually unstable. We will thus refrain from investigating further the feasibility of such a configuration.

\begin{figure}[h!]
	\centerline{\includegraphics[width=0.65\linewidth]{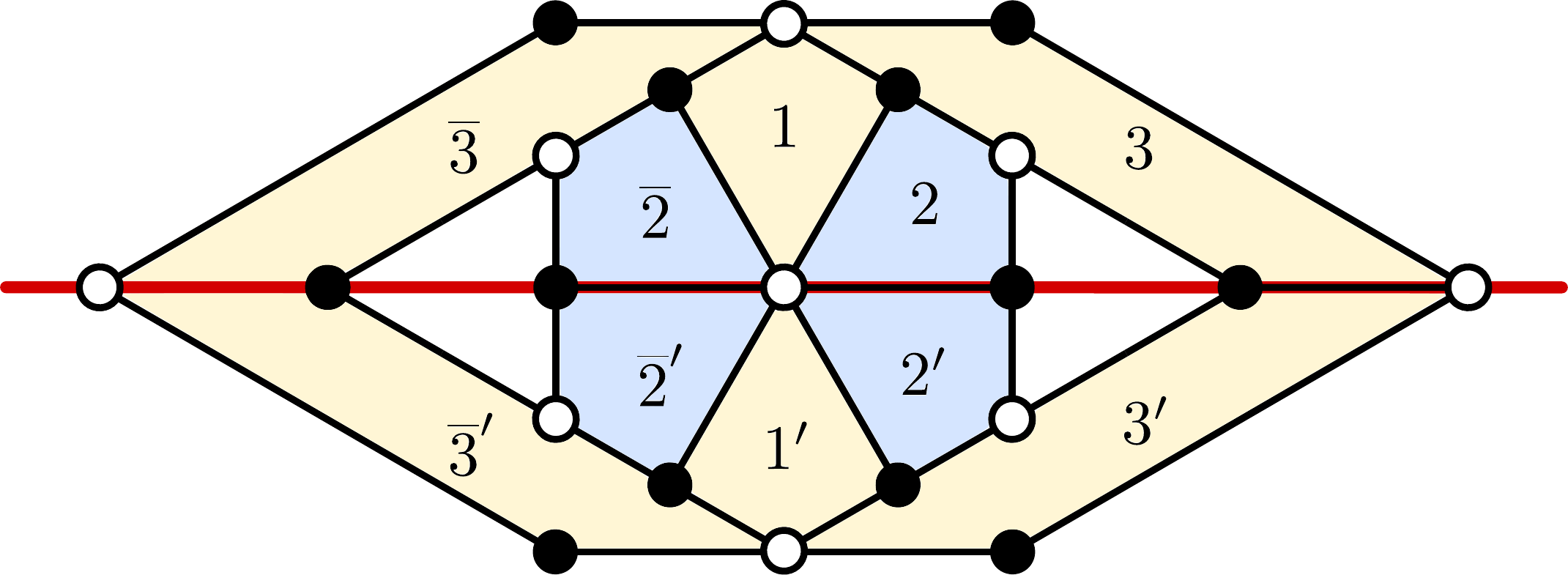}}
	\caption{A tentative sub-dimer for a twin 3-2 model where the $SU(3)$ faces have an antisymmetric flavor.}
	\label{32twin1}
\end{figure}

\begin{figure}[h!]
	\centerline{\includegraphics[width=0.5\linewidth]{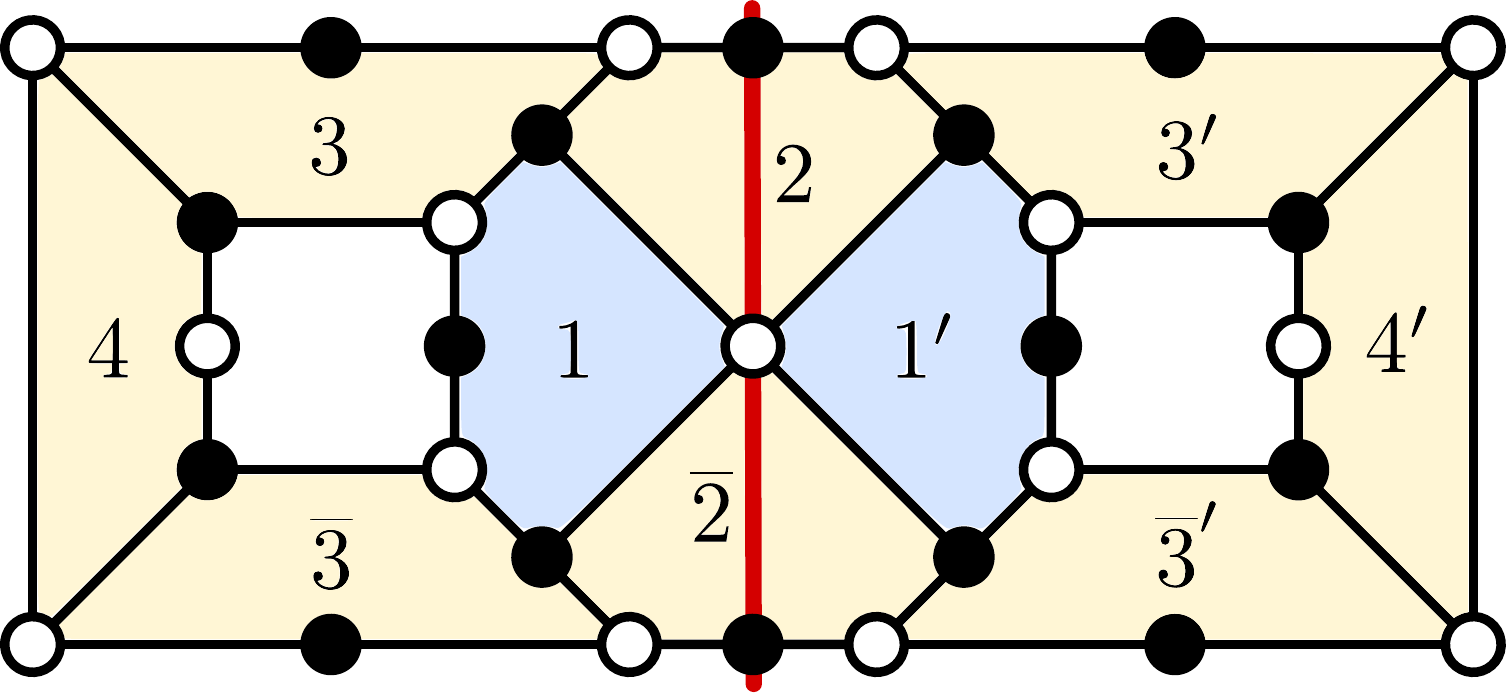}}
	\caption{A tentative sub-dimer for a twin 3-2 model where the $SU(3)$ faces share a flavor $SU(1)_4$.}
	\label{32twin2}
\end{figure}

Finally, we would like to see if it is possible to realize a twin 3-2 model on a single fixed line. Given that node 2 and its twin require a $-$ sign, in principle we have two options. Either both node 3 and its twin have an antisymmetric by ending-up on the same fixed line, or they compensate the anomaly by sharing an $SU(1)$ node 4. It is easy to draw the minimal requirements for the portion of the dimer that would translate these properties, see respectively Figures \ref{32twin1} and \ref{32twin2}. 
	
Naively, these configurations look consistent and one can find a choice of ranks satisfying the ACC. These are the following. For \fref{32twin1}, $N_3=N_{3'}=N_{\bar 3}=N_{\bar3'}=M_3+3$, $N_2=N_{2'}=N_{\bar 2}=N_{\bar2'}=M_2+2$ and $N_1=N_{1'}=M_2+M_3+1$. For \fref{32twin2}, $N_3=N_{3'}=N_{\bar 3}=N_{\bar3'}=M_3+3$,   $N_1=N_{1'}=M_1+1$,  $N_4=N_{4'}=M_1'+1$ and $N_2=N_{\bar 2}=M_1+M_1'+2$.

Assuming that in the parent theory every rank parameterizing the solutions above can be taken independently large, we observe that both situations would imply the existence of a fractional brane described by a ring of faces with equal ranks (up to the usual ${\cal O}(1)$ corrections) surrounding a hole. These are obtained by setting $M_2=0$ in \fref{32twin1}, and $M_1=0$, $M_1'=M_3$ in \fref{32twin2}. The ring-shaped would-be fractional brane is depicted in both figures by the yellow-shaded faces. As shown in Appendix \ref{dimers}, this is an inconsistent dimer. We conclude that unlike the $SU(5)$ model, there is no way to build a stable twin version of the 3-2 model on a single orientifold line.

\section{The Rise of the Octagon} 
\label{Octagon}

In \sref{su5} and \sref{32} we have shown that the only alternative for an a priori consistent realization of a DSB model which does not automatically imply the presence of an $\mathcal{N}=2$ fractional brane, and hence is potentially stable in the decoupling limit, is the twin $SU(5)$ living on a single fixed line of an orientifold. The twin $SU(5)$ model is described by the hexagonal cluster depicted in \fref{hexagon}. Now we want to understand if such cluster can be embedded in a fully consistent dimer and if such dimer can be free of $\mathcal{N}=2$ fractional branes. 

Let us first argue that in the full theory the hexagonal cluster is associated to a fractional brane. From \fref{hexagon} we see that the ACC are satisfied for $N_1=N_3=N+4$ and $N_2=N$. Namely, we are free to choose any value of $N$ while all other faces of the dimer sharing an edge with the faces of the hexagonal cluster have vanishing rank. This freedom is associated to the presence of a fractional brane. The twin $SU(5)$ is obtained for $N=1$, i.e.~a single fractional brane.
		
Now we can ask whether this fractional brane is of deformation or runaway DSB type, in the parent theory (we already know we do not want it to be of $\mathcal{N}=2$ type).  If it were a runaway DSB brane some other regions of the dimer, besides the hexagon, would be populated and the corresponding faces would have ranks with different multiples of $N$ \cite{Franco:2005zu,Bertolini:2005di}. This is the key ingredient to generate an ADS superpotential and hence a runaway behavior, and this will still be true after orientifolding. Thus a runaway DSB brane in the parent theory, if it survives the orientifold, will still be of runaway type. Populating the dimer with regular branes, the runaway sector will communicate with the twin $SU(5)$ sector, destabilizing  the vacuum. The other possibility is that the hexagonal cluster corresponds to a deformation brane in the parent theory and that it survives the orientifold projection. This has no instability in the parent theory, and thus we expect it to remain stable also upon orientifolding. 

It is known \cite{Franco:2005zu,Butti:2006hc} that deformation fractional branes are related to ZZP, see Appendix \ref{dimers}. We are looking for a dimer containing a six-valent node inside a cluster of faces. The corresponding toric diagram must contain at least 6 edges whose associated ZZP are ordered around the relevant node \cite{Gulotta:2008ef,ishii2010note}. Those edges need to be in equilibrium, and once removed the rest of the $(p,q)$-web must be in equilibrium, too. This implies that we need at least two extra ZZP in equilibrium, for a total of eight. Absence of $\mathcal{N}=2$ fractional branes in the dimer further requires that there cannot be more than one ZZP with a given winding $(p,q)$ of the unit cell. This corresponds to toric diagrams with no more than two consecutive points which are aligned on an external edge.

Since we are looking for a singularity admitting line orientifolds, we consider toric diagrams with line symmetry, either vertical/horizontal or diagonal. 

\begin{itemize}
	\item \underline{Diagonal line}
	
	From \fref{hexagon} we see that we need two antisymmetric fields, in ${\tiny \yng(1,1)_1}$ and $\overline{\tiny \yng(1,1)}_3$ representations, respectively. Even if dimer models containing the required deformation can be engineered, it turns out that 
	there is no solution to the ACC of the full dimer, as it happens for all the theories (but a very special family, which however contains $\mathcal{N}=2$ fractional branes) obtained as orientifolds of dimers with a diagonal fixed line (this is discussed in \cite{Argurio:2020dko}, where arguments connecting general features of the geometry to the solvability of the ACC are presented). Thus, such cases are excluded.
	
	\item \underline{Vertical/horizontal lines}
	
	As discussed in \sref{su5}, the freedom in choosing different charges for the two fixed lines is a crucial difference with respect to diagonal line orientifolds. In fact, it guarantees the existence of solutions to the ACC after orientifolding, exactly balancing the contribution from the different tensor fields. As discussed in \cite{Argurio:2020dko}, this is ensured by noticing that tensor fields come in pairs in the dimer, one in each of the two lines. Assigning opposite signs to the two lines grants that the two contributions cancel, yielding an anomaly free theory. If the two signs are chosen the same, the situation is the same as with diagonal lines. 
\end{itemize}

The upshot is that having vertical/horizontal lines, with opposite signs for the two orientifold lines, is the only option which can lead to viable twin $SU(5)$ models and it is what we are going to focus on in the following. 
  
The need for two tensor fields is a stringent constraint on the ZZP, and therefore on the toric diagram. In particular, it implies that two couples of ZZP must have the correct intersection number among themselves and with the fixed lines, as computed from the toric diagram, see Appendix \ref{dimers} and \cite{Argurio:2020dko}. 

\begin{figure}[!!!t]
	\begin{center}
		\includegraphics[width=0.20 \textwidth]{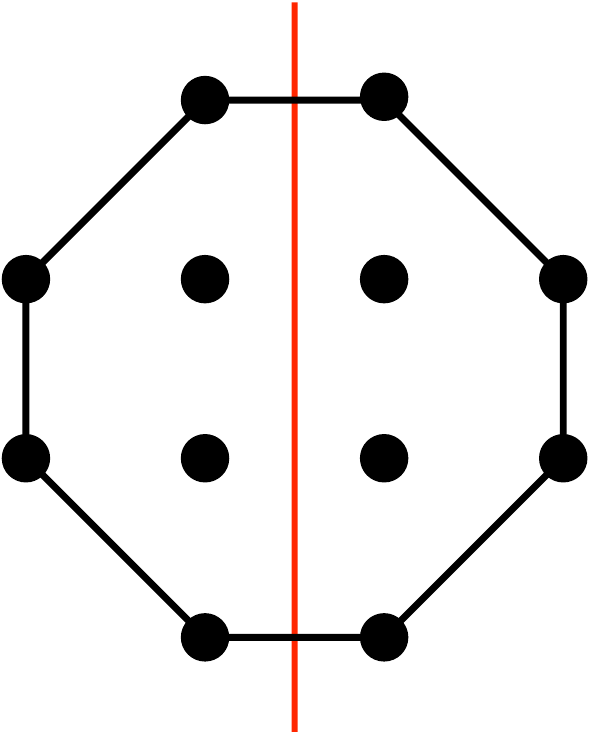}
	\end{center}
	\vspace{-.3cm}
	\caption{\small The toric diagram of the Octagon singularity.
		\label{fig:toricdiagram}}
\end{figure}

Remarkably, the aforementioned necessary conditions provide substantial guidance for where to look for a model that works, as we now explain. The simplest example of a toric diagram with the required eight ZZP, with the correct intersection numbers, no $\mathcal{N}=2$ fractional branes and the necessary horizontal symmetry is the toric diagram depicted in \fref{fig:toricdiagram}, that we dub the Octagon. 

\begin{figure}[!!!t]
\begin{center}
\includegraphics[width=0.38 \textwidth]{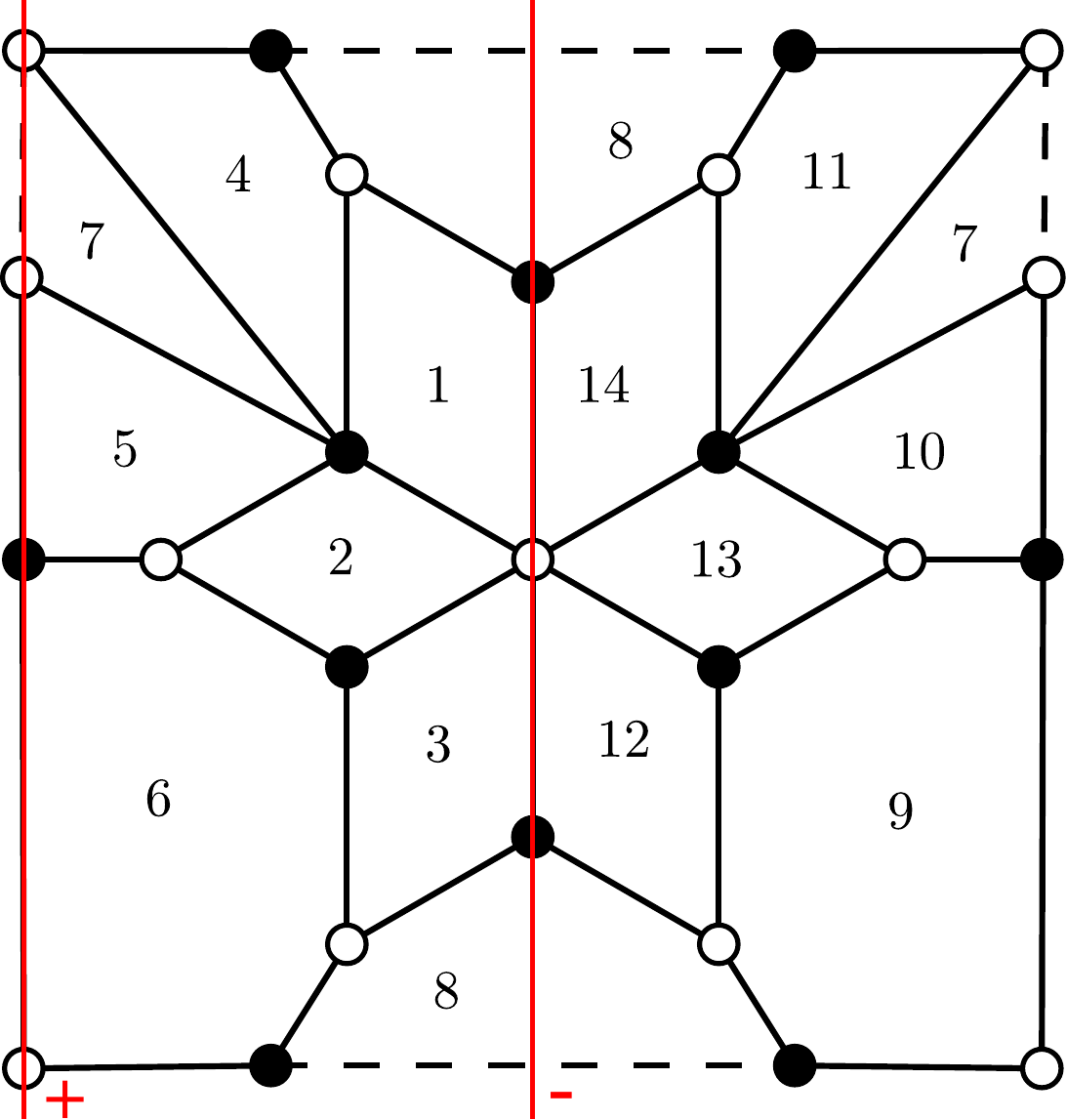}
\end{center}
\vspace{-.3cm}
\caption{\small The (unit cell of the) dimer describing the symmetric phase of the Octagon. Orientifold lines are in red. Each orientifold line has a sign associated to it, which in this case needs to be opposite one another. 
\label{fig:dimer}}
\end{figure}

Using standard techniques one can associate a dimer to a toric diagram, one for each different toric phase \cite{Franco:2005rj,Franco:2005sm}. A generic toric phase does not display the symmetry required to perform the orientifold projection. In the present case, however, one can find a symmetric toric phase where the vertical fixed lines are manifest and which realizes the twin $SU(5)$ model as described above. The corresponding dimer is depicted in \fref{fig:dimer}, where the hexagonal cluster is described by the white dot in the center of the unit cell. A quick and direct way to check that the dimer in \Cref{fig:dimer} does correspond to the toric diagram in \Cref{fig:toricdiagram} is by the Fast Forward Algorithm \cite{Franco:2005rj}, as detailed in Appendix \ref{dimers}.

Let us look at the orientifold gauge theory more closely. The orientifold projection identifies faces $(1,\ldots,6)$ with faces $(14,\ldots,9)$ while faces 7 and 8 are self-identified. Hence, D-branes at such orientifold singularity are described by a matter coupled supersymmetric gauge theory with six $SU$ factors, one $SO$ and one $USp$ factors. The twin $SU(5)$ model is given by the rank assignment $SU(5)_1 \times SU(1)_2 \times SU(5)_3$ with all other faces being empty but face 7 which is a decoupled pure SYM with gauge group $SO(5)$ and hence confines on its own. ACC and self-consistency of such rank assignment follow the general discussion in \sref{su5}. 

More details on the Octagon and its physical properties can be found in \cite{Argurio:2020dkg}. Here it suffices to say that this model represents a concrete example of an orientifold singularity which allows DSB by a D-brane bound state which is free of any known instability, in particular the $\mathcal{N}=2$ fractional brane decay channel or the runaway behavior typical of DSB branes. The absence of $\mathcal{N}=2$ fractional branes is clear from the toric diagram in \fref{fig:toricdiagram}, which does not have internal points on boundary edges. This model therefore provides a realization (the first, to our knowledge) of stable DSB with D-branes at CY singularities and suggests for an extension of the string theory landscape as it is currently known. 

\vskip 10pt

The Octagon emerges as the simplest possible dimer having all required properties to admit, upon orientifolding, stable DSB D-brane configurations. One might ask whether less minimal models exist which share the same properties. We do not have an answer to this question, yet. Still, dimer techniques have (once again) proven to be a very powerful tool to provide a direct link between geometry and gauge theories dynamics, both in finding no-go theorems, like the one presented in \cite{Argurio:2019eqb} or the connection between minimal $SU(5)$ and 3-2 models and the presence of $\mathcal{N}=2$ fractional branes established in this paper, as well as in unveiling concrete ways to evade them. Therefore, we cannot exclude that further surprises are possible and generalizations of the Octagon model will eventually be found.

\acknowledgments
We are grateful to I\~naki Garc\'ia-Etxebarr\'ia, Ander Retolaza and Angel Uranga for useful discussions and comments.  R.A., A.P. and E.G.-V. acknowledge support by IISN-Belgium (convention 4.4503.15) and by the F.R.S.-FNRS under the ``Excellence of Science" EOS be.h project n.~30820817, M.B. and S.M. by the MIUR PRIN Contract 2015 MP2CX4 ``Non-perturbative Aspects Of Gauge Theories And Strings" and by INFN Iniziativa Specifica ST\&FI. E.G.-V. was also partially supported by the ERC Advanced Grant ``High-Spin-Grav". The research of S.F. was supported by the U.S. National Science Foundation grants PHY-1820721 and DMS-1854179. R.A. is a Research Director and A.P. is a FRIA grantee of the F.R.S.-FNRS (Belgium).

\appendix

\section{Dimers and Orientifolds}
\label{dimers}

Dimer diagrams \cite{Franco:2005rj,Franco:2005sm} are exceptional tools in describing the gauge theories on D3-branes probing toric CY  singularities. In this appendix we review them, along with some of their combinatorial tools and their relation to the toric diagram of the underlying geometry. We also review how orientifolds are added to the game.

\begin{figure}[h!]
	\centering
	\begin{subfigure}[t]{0.28\textwidth }
		\begin{center} 
			\includegraphics[width=\textwidth]{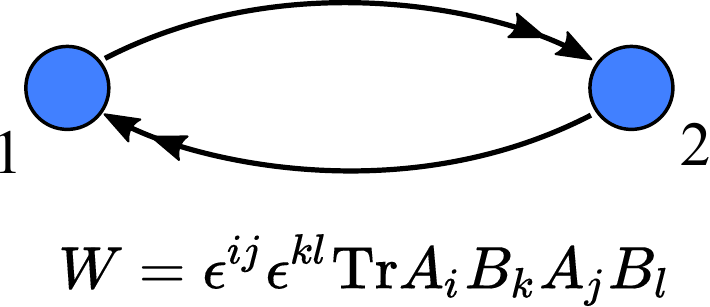}
			\caption{}
			\label{Fig:ConifoldQuiver}
		\end{center}
	\end{subfigure} \hspace{15mm}
	\begin{subfigure}[t]{0.30\textwidth } 
		\begin{center} 
			\includegraphics[width=\textwidth]{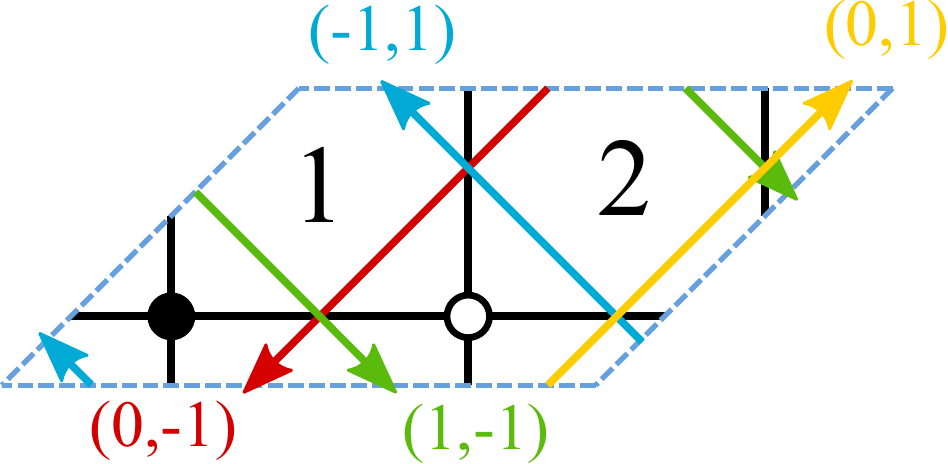}
			\caption{}
			\label{Fig:ConifoldZigZags}
		\end{center}
	\end{subfigure} \hspace{15mm}
	\begin{subfigure}[t]{0.18\textwidth } 
		\begin{center} 
			\includegraphics[width=\textwidth]{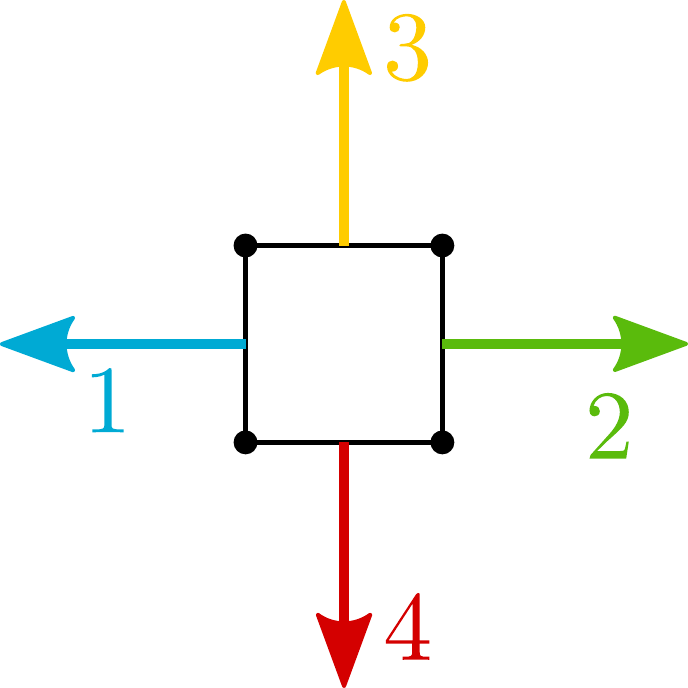}
			\caption{}
			\label{Fig:ToricDiagramConifold}
		\end{center}
	\end{subfigure}
	\caption{a) Quiver diagram of the conifold with superpotential. b) Dimer diagram of the conifold with ZZP and their holonomies. c) Toric diagram with schematic web diagram corresponding to ZZP up to an $SL(2,\mathbb{Z})$ transformation.}\label{Fig:Conifold} 
\end{figure}

A set of D3-branes probing a toric CY singularity hosts an $\mathcal{N}=1$ gauge theory given by a bunch of $SU(N)$ gauge groups and bifundamental chiral fields appearing exactly in two monomials with opposite sign in the superpotential. This constrained structure allows their embedding on a bipartite tiling of $\mathbb{T}^2$ called Dimer diagram. Faces in the tiling correspond to $SU(N)$ gauge groups, edges are bifundamental fields and nodes are superpotential terms. The orientation of fields is set by going clockwise around black nodes and counter-clockwise around white nodes. Furthermore, we adopt a convention in which the tail and head of an arrow correspond to the fundamental and antifundamental representations, respectively. Superpotential terms are read concatenating fields around nodes, their signs given by the color of the node, 
$+$ for white nodes and $-$ for black nodes. A prototypical example is the conifold, shown in \Cref{Fig:Conifold}. 

Several algorithms allow to find the toric diagram of the underlying geometry given a dimer diagram. The simplest one uses the Kasteleyn matrix $K$ \cite{Franco:2005rj,kenyon2003introduction,Kenyon:2003uj,Hanany:2005ve},  whose determinant is the Newton polynomial of the toric diagram. A related, more indirect but more graphical method uses {\it zig-zag paths} (ZZP), oriented paths in the graph that go along edges and turn maximally right at white nodes and maximally left at black nodes. These paths have an holonomy around $\mathbb{T}^2$ precisely corresponding to the $(p,q)$ labels of the web diagram dual to the toric diagram. One can then use them to find the toric diagram, see \Cref{Fig:Conifold}.

\begin{figure}[h!]
	\centering
	\begin{subfigure}[t]{0.35\textwidth }
		\begin{center} 
			\includegraphics[width=\textwidth]{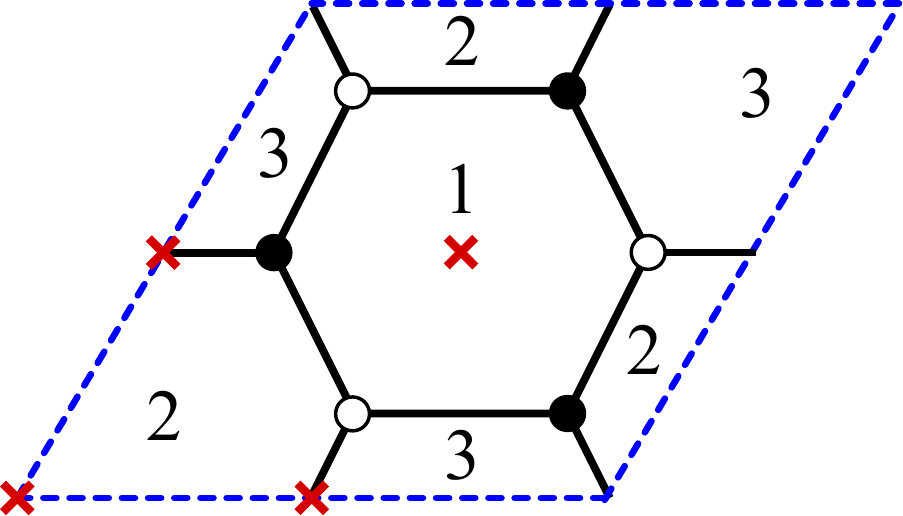}
			\caption{}
			\label{Fig:OrientifoldPoints}
		\end{center}
	\end{subfigure} \hspace{15mm}
	\begin{subfigure}[t]{0.20\textwidth } 
		\begin{center} 
			\includegraphics[width=\textwidth]{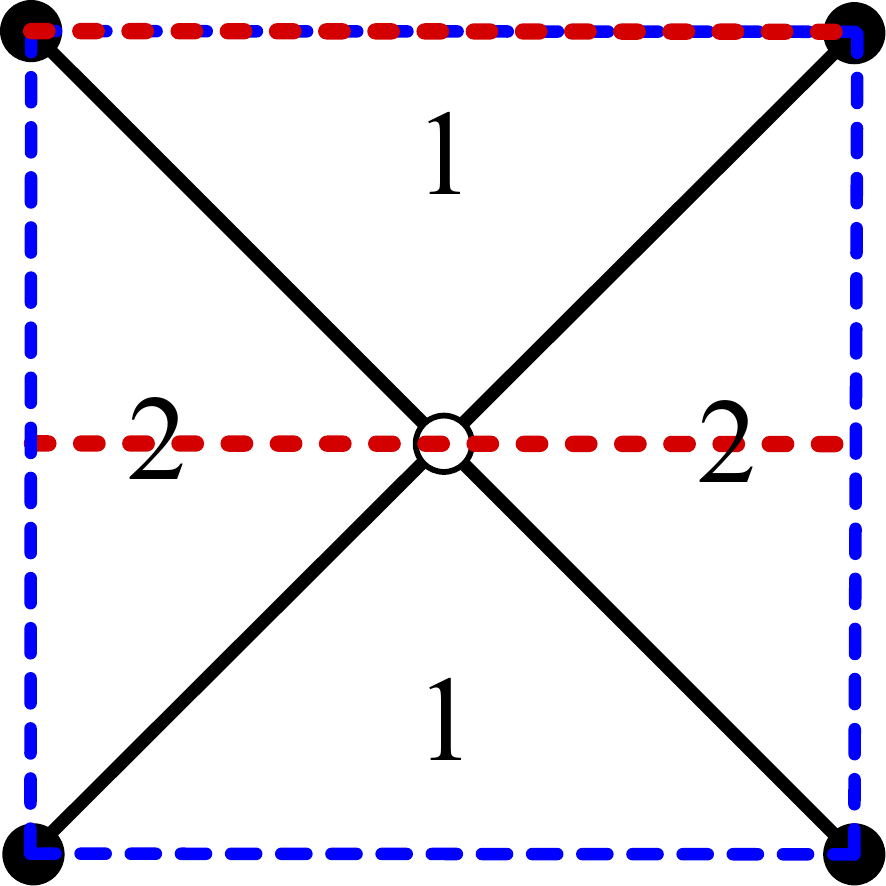}
			\caption{}
			\label{Fig:OrientifoldLine}
		\end{center}
	\end{subfigure} 
	\caption{a) Orientifold of $\mathbb{C}^3/\mathbb{Z}_3$ with fixed points. b) Orientifold of the conifold with fixed lines. Note that the unit cell has been transformed with respect to \Cref{Fig:ConifoldZigZags}.}\label{Fig:Orientifold} 
\end{figure}

Whenever one is probing these geometries with D3-branes, one may add orientifold planes. The resulting projection on the gauge theory is described as a geometric projection on the dimer \cite{Franco:2007ii} with either 4 fixed points or (1 or 2) fixed lines, see \Cref{Fig:Orientifold}. The different fixed loci are assigned different signs giving rise to different theories. 

\paragraph{Fixed Points.} 
The unit cell may always be chosen such that the fixed points lie at the origin, middle of two sides and center of the cell, as in \Cref{Fig:OrientifoldPoints}. Their signs obey the following rule: their product must be $+$ or $-$ if the number of nodes is a multiple of $4$ or $2$. Self-identified faces represent $SO(N)$ or $USp(N)$ groups for $+$ and $-$ sign, respectively.  Otherwise, they combine with their images and represent an $SU(N)$ group. Likewise, self-identified matter fields represent symmetric or antisymmetric fields for $+$ and $-$ sign. Otherwise they are bifundamental matter. An example is shown in \Cref{Fig:OrientifoldPoints}. Taking $(+,+,-,+)$ signs running counter-clockwise and starting at the origin, one gets an $USp(N_1)_1 \times SU(N_2)_2$ theory with matter fields
\begin{equation}
3 \symm_2 + 3 (\square_1, \overline\square_2)\ ,
\end{equation}
which lead to a non-anomalous theory for $N_1=N_2+4$.
\paragraph{Fixed Lines.} Depending on the symmetry of the unit cell there may be one diagonal line or two horizontal lines. Their signs are not restricted. The relevant $\mathbb{Z}_2$ symmetry shows up on the toric diagram \cite{Retolaza:2016alb}. An example is shown in \Cref{Fig:OrientifoldLine} with two horizontal lines, which describes a gauge theory with two gauge groups, $SO(N)$ or $USp(N)$ depending on the chosen signs, with two bifundamentals. 

\subsection{Holes in the Dimer and Zig-Zag Paths}

In the following we present an argument forbidding the presence of holes of reduced rank inside a specific sub-dimer which appears in different twin models. We rely on ZZP techniques for anomaly cancellation developed in \cite{Argurio:2020dko,Butti:2006hc}. One associates a value $v_i$ to every ZZP in the dimer and then assigns an arbitrary rank to a given face in the dimer. The remaining ranks are set by requiring that the rank differences between two adjacent faces $m,n$ obey $N_m - N_n = v_i-v_j$  where $i,j$ are the ZZP separating them.

\begin{figure}[h!]
	\centering
	\begin{subfigure}[t]{0.29\textwidth }
		\begin{center} 
			\includegraphics[width=\textwidth]{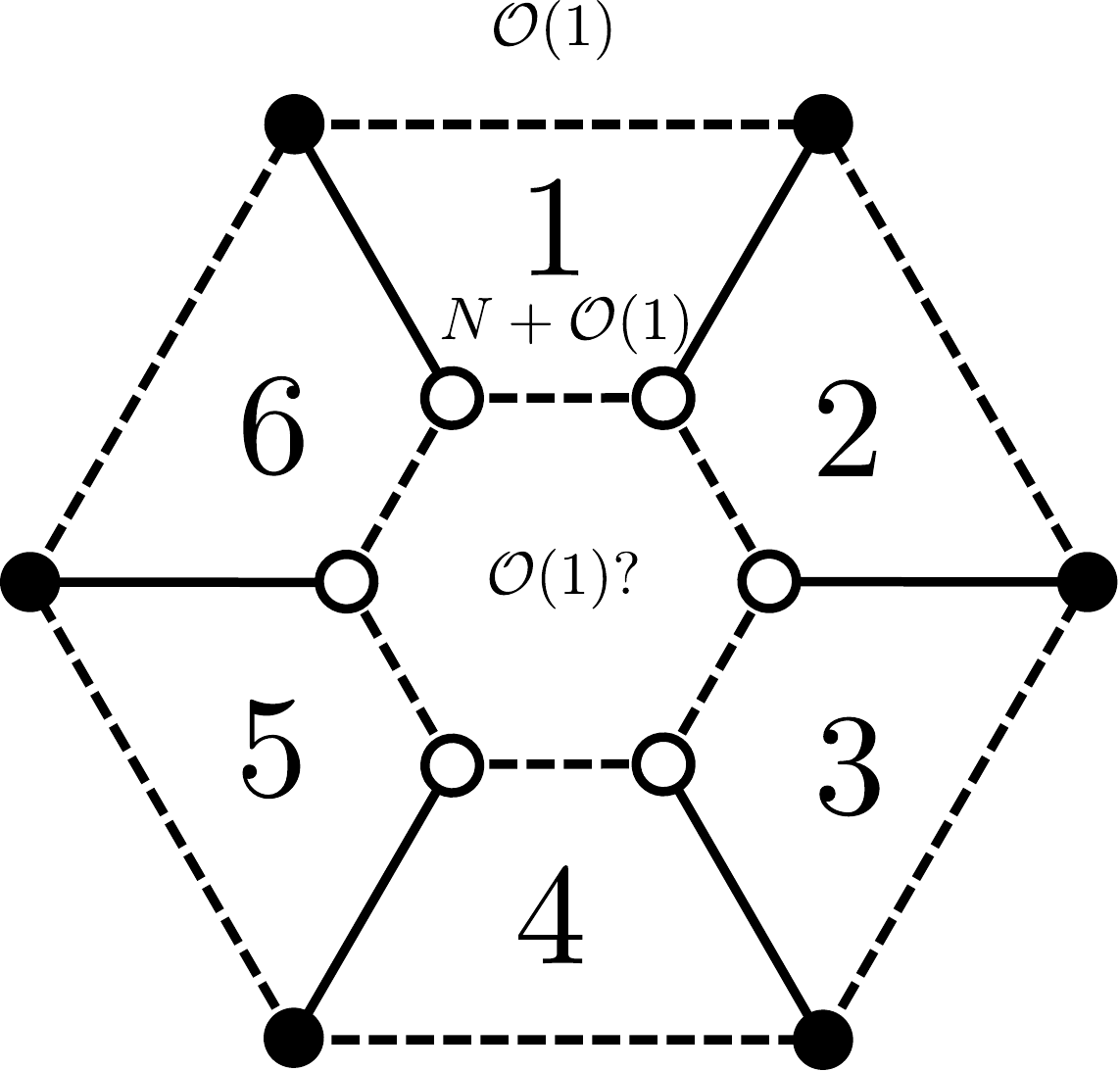}
			\caption{}
			\label{Fig:Hole}
		\end{center}
	\end{subfigure} \hspace{12mm}
	\begin{subfigure}[t]{0.24\textwidth } 
		\begin{center} 
			\includegraphics[width=\textwidth]{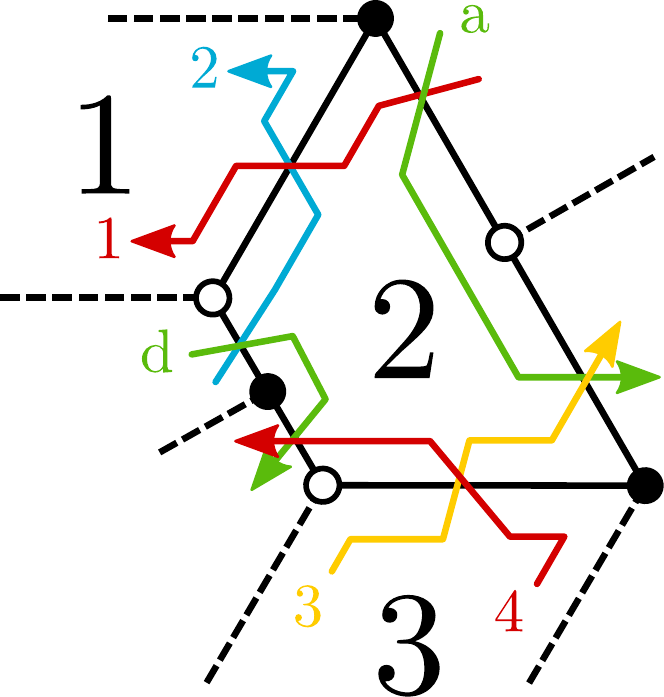}
			\caption{}
			\label{Fig:HoleZigZag6}
		\end{center}
	\end{subfigure} 
	\caption{(a) Generic ring of rank $N + \mathcal{O}(1)$ surrounded by faces of rank $\mathcal{O}(1)$ with a hole of rank $\mathcal{O}(1)$. (b) Face 2 edges with zig-zag paths.}\label{Fig:Holes} 
\end{figure}

Consider a ring-shaped sub-dimer of rank $N + \mathcal{O}(1)$. We assume that as we go along it, from one of its faces to another, we only cross edges with identical orientation, see \Cref{Fig:Hole}. We now show that the region inside the ring, the ``hole", is inconsistent if of reduced rank. 

Consider a face of the ring, as face 2 in \Cref{Fig:HoleZigZag6}. The intersections between the ZZP 1, 2, 3 and 4 yield
\begin{equation}
N_1-N_2 = v_1-v_2 \sim  0 \, , \quad N_2-N_3 = v_4-v_3 \sim 0 \, ,  \label{Eq:0mod4}
\end{equation}
where $\sim$ means ``equal up to $\mathcal{O}(1)$". Since the hole is supposed to be of rank $\mathcal{O}(1)$, the intersections with Zig-Zags that separate it from the ring give
\begin{equation}
N \sim v_2 - v_d, \quad -N \sim v_d - v_4, \quad \Rightarrow \quad v_2 \sim v_4 \, . \label{Eq:v2}
\end{equation}
Changing the number of edges between face 2 and the hole can only be done by adding/removing pairs of edges and will not change the fact that
\begin{equation}
v_1 \sim v_2 \sim v_3 \sim v_4 \qquad \text{ and } \qquad v_d \sim v_1 - N \, ,
\end{equation}
where $v_d$ is understood as any ZZP that comes with the pair of edges added between the hole and face 2. One can repeat the reasoning for every face of the ring and find that its internal edges will be always produced by ZZP $\sim v_1$. This is in contradiction with the presence of ZZP $v_d \sim v_1 - N$ since there are only ZZP $\sim v_1$ entering the hole. It implies that $v_d$ is circular or not present. The first option is forbidden in dimer models and the second spoils the presence of the hole itself. Hence the presence of an anomaly-free hole inside such a ring is inconsistent.

As a comment, let us notice that to reach this conclusion we did not assume anything about the exterior of the ring. If one does not look at the hole but asks that the exterior has a reduced rank, it implies that ZZP $v_a$ on its border, see \Cref{Fig:HoleZigZag6}, will satisfy
\begin{equation}
v_a \sim v_1 + N \sim v_3 + N \, , \label{Eq:v1}
\end{equation}
and thus we recover the result of \Cref{Eq:v2} using \Cref{Eq:0mod4}. Again, it can be shown that this result does not depend on the number of edges in contact with the exterior of the ring. The cluster (hexagonal or otherwise) is now viable only with ranks $N + \mathcal{O}(1)$, because it is made only of ZZP $\sim v_1$.

\subsection{The Octagon and its Symmetric Phase}

As discussed in \cite{kenyon2003introduction,Kenyon:2003uj,Hanany:2005ve}, to any dimer model one can associate a weighted, signed adjacency matrix, known as the Kasteleyn matrix, whose determinant is the characteristic polynomial of the dimer model from which one can extract the toric data. This procedure is known as the Fast Forward Algorithm and is reviewed in \cite{Franco:2005rj}.

\begin{figure}[h!]
	\centering
	\includegraphics[width=0.5\textwidth]{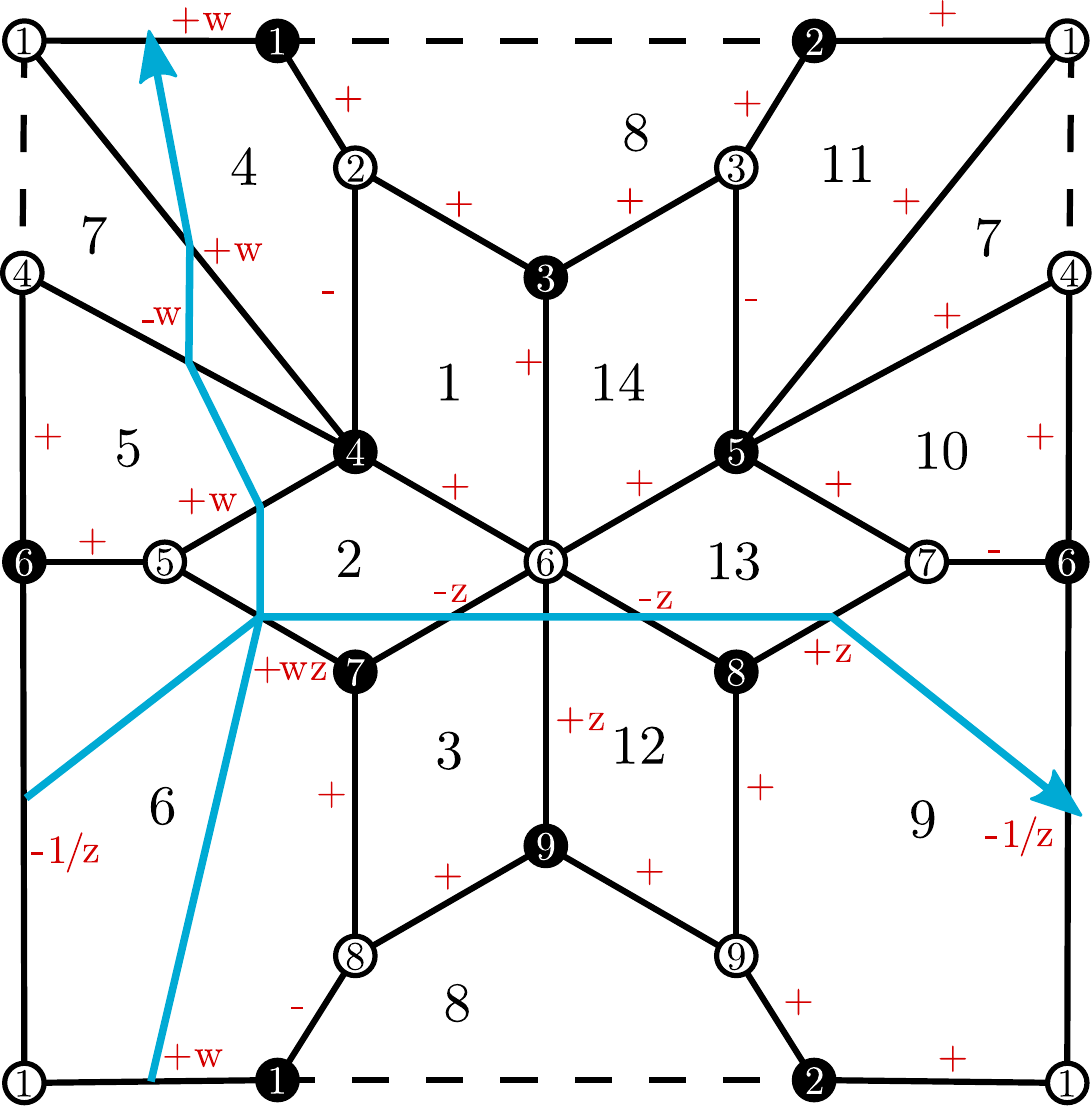}
	\caption{Dimer diagram of the Octagon with weights (in red) for building the Kasteleyn matrix. White and black nodes have been numbered. Two fundamental paths are shown in blue.}\label{Fig:octagonKast2} 
\end{figure}

To obtain the Kasteleyn matrix one assigns a sign to every edge such that for every face in the dimer the product of signs is $+1$ if its number of edges is $2 \, \text{mod} \, 4$ and $-1$ if its number of edges is $0 \, \text{mod} \, 4$. One then constructs two closed oriented (gauge invariant) paths $\gamma_w, \gamma_z$ with holonomy $(0,1)$ and $(1,0)$. Every edge crossed by these paths is multiplied by $w$ or $1/w$, depending on the relative orientation (respectively by $z$ or $1/z$). The resulting graph for the Octagon is shown in \Cref{Fig:octagonKast2}.

The adjacency matrix of the graph with such weights is the Kasteleyn Matrix and, for the Octagon, it reads

\beq
K = \left( 
\begin{array}{r|ccccccccc}
& \ \ \ 1 \ \ & \ \ 2 \ \ & \ \ 3 \ \ & \ \ 4 \ \ & \ \ 5 \ \ & \ \ 6 \ \ & \ \ 7 \ \ & \ \  8 \ \ & \ \ 9 \ \ \\ \hline
1 \ \ & \ w & 1 & 0 & w & 1 & -\frac{1}{z} & 0 & 0 & 0 \\ 
2 \ \ & \ 1 & 0 & 1 & -1 & 0 & 0 & 0 & 0 & 0 \\ 
3 \ \ & \ 0 & 1 & 1 & 0 & -1 & 0 & 0 & 0 & 0 \\ 
4 \ \ & \ 0 & 0 & 0 & -w & 1 & 1 & 0 & 0 & 0 \\ 
5 \ \ & \ 0 & 0 & 0 & w & 0 & 1 & w z & 0 & 0 \\ 
6 \ \ & \ 0 & 0 & 1 & 1 & 1 & 0 & -z & -z & z \\ 
7 \ \ & \ 0 & 0 & 0 & 0 & 1 & -1 & 0 & z & 0 \\ 
8 \ \ & \ -1 & 0 & 0 & 0 & 0 & 0 & 1 & 0 & 1 \\ 
9 \ \ & \ 0 & 1 & 0 & 0 & 0 & 0 & 0 & 1 & 1 
\end{array}
\right)
\eeq
where rows and columns correspond to white and black nodes in the dimer, respectively. Its determinant is
\begin{equation}
\text{det}(K) = w^3 z^2+w^3 z+w^2 z^3-24 w^2 z^2+26 w^2 z-w^2+w z^3+24 w z^2+26 w z+w-z^2+z~.
\end{equation}
One may compute the Newton Polygon of the above expression and it should correspond to the toric diagram of the dimer one is dealing with \cite{Hanany:2005ve}. For every monomial $a ~w^b z^c$ one draws a point in a $2d$ lattice with coordinates $(b,c)$. As expected, one obtains the toric diagram depicted in \fref{fig:toricdiagram}. Nicely, there is a single perfect matching for each of its external points, thus ensuring that the dimer meets a necessary condition of minimality.

\section{ACC for 3-2 Quivers} 
\label{appendix_ACC_32}

Not all of the quivers presented in \fref{quiver_32_1} are free of anomalies when $N_i \neq 0$. In this appendix we check this explicitly. Our calculations also motivate the choice of the antisymmetric tensor $\wb A_2$ to satisfy the ACC. Below we summarize the ACC for each of these models. For completeness, we added here as different cases also the two models where node 1 and 4 are identified.

\begin{itemize}
	\item \underline{$SO(N_1 +1)\times USp(N_2 +2) \times SU(N_3 +3)$ with ${\tiny {\yng(1,1)}_3}$}:
	\begin{eqnarray}
	\mbox{{\bf Node 3:}}	\ \  (N_3 +3 -4) - (N_1 +1) + (N_2 +2) &=& 0 \, .
	\end{eqnarray}
	
	\item \underline{$SO(N_1 +1)\times SU(N_2 +2) \times SU(N_3 +3)$ with ${\tiny {\yng(1,1)}_3}$}:
\beq
\begin{array}{lrl}
\mbox{{\bf Node 2:}}	\ \ & -(N_2 + 2-4) + (N_1 + 1) - (N_3 +3) & = 0 \ ,\\
\mbox{{\bf Node 3:}}	\ \ &	(N_3 +3 -4) - (N_1 +1) + (N_2 +2) & = 0\ .
\end{array}	
\eeq
	Note that the choice of conjugate representation for the antisymmetric tensor of $SU(N_2 +2)$ is fixed by the first equation, in order to satisfy it when all $N_i = 0$. \
	
For these two first models, the ACC reduce to
	\begin{eqnarray}
	N_1 = N_2 +N_3 \, .
	\end{eqnarray}
	
	\item \underline{$SO(N_1 +1)\times USp(N_2 +2) \times SU(N_3 +3)\times SO(N_4 +1)$}:
\beq
	\mbox{{\bf Node 3:}}	\ \ - (N_1 +1) + (N_2 +2) - (N_4 +1) = 0 \, .
\eeq
In this case, $N_3$ is not constrained by the ACC, which can be rewritten as
\beq
	N_2 = N_1 + N_4\, .
\eeq
	
	\item \underline{$SO(N_1 +1)\times SU(N_2 +2) \times SU(N_3 +3)\times SO(N_4 +1)$}:
\beq
\begin{array}{lrl}
\mbox{{\bf Node 2:}}	\ \ & 	-(N_2 + 2-4) + (N_1 + 1) - (N_3 +3) & = 0 \ ,\\
\mbox{{\bf Node 3:}}	\ \ & 	- (N_1 +1) + (N_2 +2) - (N_4 +1)  & = 0\ .
\end{array}	
\eeq
	This translates to the two conditions
\beq
\begin{array}{rl}
	N_1 = & N_2 +N_3\, , \\
	N_2 = & N_1 + N_4\, ,
\end{array}	
\eeq
	implying $N_3=-N_4$. This in turn sets $N_3=N_4=0$, since all $N_i$ must be positive and potentially large.
	In principle this issue does not rule out the possible engineering of these models, since the corresponding dimers might give rise to additional gauge groups and fields when regular D3-branes are added, in a way that anomalies are cancelled. Assuming that at least some fractional branes are needed in order to turn on all the ranks of the 3-2 model (i.e.~even for $N_i=0$), then such models are excluded. 
	
	\item \underline{$SO(N_1 +1)\times USp(N_2 +2) \times SU(N_3 +3)$ with $2(\tiny \overline {\yng(1)}_3 , \tiny {\yng(1)}_1)$}:
\beq
	- 2 (N_1 +1) + (N_2 +2)  = 0 \, ,
\eeq
which is simply
\beq
	N_2 = 2 N_1 \, .
\eeq
	
	\item \underline{$SO(N_1 +1)\times SU(N_2 +2) \times SU(N_3 +3)$ with $2(\tiny \overline {\yng(1)}_3 , \tiny {\yng(1)}_1)$}:
\beq
\begin{array}{lrl}
\mbox{{\bf Node 2:}}	\ \ & -  (N_2 +2-4) + (N_1 +1)  -(N_3+3) & = 0 \ ,\\
\mbox{{\bf Node 3:}}	\ \ & - 2 (N_1 +1) + (N_2 +2)  & = 0\ .
\end{array}
\eeq
	This can be simplified into
\beq
\begin{array}{rl}
	N_2 = & 2 N_1 \ ,\\
	N_3 = & - N_1\ ,
\end{array}
\eeq
which has no solution beyond $N_i=0$ in the absence of additional ingredients coming from the full dimer.
\end{itemize}

The results of this appendix can be summarized in the following table:
\begin{center}
	\begin{tabular}{c|c}
		Gauge groups & ACC \\
		\hline
		$SO(N_1 +1)\times USp(N_2 +2) \times SU(N_3 +3)$ with ${\tiny {\yng(1,1)}_3}$ & \cmark \\
		$SO(N_1 +1)\times SU(N_2 +2) \times SU(N_3 +3)$ with ${\tiny {\yng(1,1)}_3}$ & \cmark \\
		$SO(N_1 +1)\times USp(N_2 +2) \times SU(N_3 +3)\times SO(N_4 +4)$ &  \cmark\\
		$SO(N_1 +1)\times SU(N_2 +2) \times SU(N_3 +3)\times SO(N_4 +4)$ & \xmark \\
		$SO(N_1 +1)\times USp(N_2 +2) \times SU(N_3 +3)$  with $2(\tiny \overline {\yng(1)}_3 , \tiny {\yng(1)}_1)$& \cmark\\
		$SO(N_1 +1)\times SU(N_2 +2) \times SU(N_3 +3)$  with $2(\tiny \overline {\yng(1)}_3 , \tiny {\yng(1)}_1)$& \xmark 
	\end{tabular}
\end{center}

\bibliographystyle{JHEP}
\bibliography{mybib}

\end{document}